\PassOptionsToPackage{numbers,sort&compress}{natbib}
\documentclass{article}

\usepackage[eandd,final]{neurips_2026}

\usepackage[utf8]{inputenc}
\usepackage[T1]{fontenc}
\usepackage{booktabs}
\usepackage{amssymb}
\usepackage{microtype}
\usepackage{xcolor}
\usepackage{hyperref}
\usepackage{graphicx}

\usepackage{colortbl}
\usepackage{xurl}
\usepackage{listings}
\usepackage{enumitem}
\usepackage{xspace}

\hypersetup{
  colorlinks,
  linkcolor={red!50!black},
  citecolor={blue!50!black},
  urlcolor={blue!80!black}
}

\newcommand{\tool}[0]{VeruSAGE\xspace}
\newcommand{\autoverus}[0]{AutoVerus\xspace}

\definecolor{javared}{rgb}{0.6,0,0}
\definecolor{javapurple}{rgb}{0.5,0,0.35}
\definecolor{javadocblue}{rgb}{0.25,0.35,0.75}

\definecolor{rankone}{rgb}{0.72, 0.90, 0.78}
\definecolor{ranktwo}{rgb}{1, 1, 1}
\definecolor{rankthree}{rgb}{1, 1, 1}

\definecolor{rankfour}{rgb}{1.0, 0.8, 0.8}

\lstdefinestyle{mystyle}{
      language=C++,
        basicstyle=\scriptsize\ttfamily,
        keywordstyle=\color{javapurple}\bfseries,
        stringstyle=\color{javared},
        commentstyle=\color{javadocblue},
        morecomment=[s][\color{javadocblue}]{/**}{*/},
        numbers=left,
        breaklines=true,
        numberstyle=\tiny\color{black},
        stepnumber=1,
        numbersep=5pt,
        tabsize=2,
        showspaces=false,
        showstringspaces=false,
        morekeywords={foreach, uint64_t, uint16_t},
        classoffset=0,
        xleftmargin=1.8em,
        escapeinside={(*@}{@*)},
        moredelim=*[is][\color{red}]{[[[}{]]]},
        captionpos=b,
}
\newcommand{\codeIn}[1]{{\small\texttt{#1}}}
\newcommand{\llm}{{LLM}\xspace} 

\newcommand{\MyPara}[1]{\vspace{.1em}\noindent\textbf{\textit{#1}}~}

\newcommand{\tablestyle}{\centering\footnotesize\setlength{\tabcolsep}{4pt}}

\newcommand{\captionfonts}{\small}

\makeatletter
\long\def\@makecaption#1#2{%
  \vskip\abovecaptionskip
  \sbox\@tempboxa{{\captionfonts #1: #2}}%
  \ifdim \wd\@tempboxa >\hsize
    {\captionfonts #1: #2\par}
  \else
    \hbox to\hsize{\hfil\box\@tempboxa\hfil}%
  \fi
  \vskip\belowcaptionskip}
\makeatother

\newcommand{\tasks}{849\xspace}
\newcommand{\noltasks}{460\xspace}
\newcommand{\sfourfivetasks}{688\xspace}
\newcommand{\projects}{eight\xspace}
\newcommand{\verusys}{VeruSAGE\xspace}
\newcommand{\bench}{VeruSAGE-Bench\xspace}
\newcommand{\benchhard}{VeruSAGE-Bench-Hard\xspace}
\newcommand{\benchurl}{\url{https://github.com/microsoft/verus-proof-synthesis}}

\lstdefinestyle{ruststyle}{
    morekeywords={
        as, break, const, continue, crate, else, enum, extern, false, fn, for, if, impl, in, let, loop, match, mod, move, mut, pub, ref, return, self, Self, static, struct, super, trait, true, type, unsafe, use, where, while, dyn, abstract, alignof, become, box, do, final, macro, offsetof, override, priv, proc, pure, sizeof, typeof, unsized, virtual, yield, async, await, try, assert
    },
    basicstyle=\small\ttfamily,
    sensitive=true,
    morecomment=[l]{//},
    morecomment=[s]{/*}{*/},
    morestring=[b]",
    morestring=[b]{'},
    keywordstyle=\color{codepurple},
    commentstyle=\color{codegray}\itshape,
    stringstyle=\color{blue},
    identifierstyle=\color{black},
    ndkeywordstyle=\color{purple}\bfseries,
    basicstyle=\ttfamily\scriptsize,
    showstringspaces=false,
    tabsize=4,
    breaklines=true,
    breakatwhitespace=false,
    showtabs=false,
    showspaces=false,
    showstringspaces=false,
    numbers=left,
    numberstyle=\tiny\color{gray},
}

\lstdefinelanguage{Verus}{
    style=ruststyle,
    basicstyle=\scriptsize\ttfamily,
    morekeywords=[2]{ requires, ensures, invariant, spec, proof, decreases},
    keywordstyle=[2]\color{red}
}

\lstdefinestyle{paperlisting}{
    basicstyle=\fontsize{6.3}{7.4}\selectfont\ttfamily,
    numberstyle=\fontsize{5.5}{7.4}\selectfont\color{gray}, numbersep=4pt,
    xleftmargin=1.2em, frame=tb, framerule=0.4pt, rulecolor=\color{black!50},
    framesep=2pt, aboveskip=0pt, belowskip=2pt,
    columns=fullflexible, keepspaces=true, breaklines=false,
}
\newcommand{\Redundant}{\makebox[0pt][l]{\color{red!10}\rule[-2.3pt]{\linewidth}{7.6pt}}}
\newcommand{\Added}{\makebox[0pt][l]{\color{green!15}\rule[-2.3pt]{\linewidth}{7.6pt}}}

\definecolor{codegray}{rgb}{0.5,0.5,0.5}
\definecolor{codepurple}{rgb}{0.58,0,0.82}

\newcommand{\sfourfive}{Sonnet 4.5\xspace}
\newcommand{\sfour}{Sonnet 4\xspace}
\newcommand{\opusfourfive}{Opus 4.5\xspace}
\newcommand{\gfive}{GPT-5\xspace}
\newcommand{\gmini}{o4-mini\xspace}
\newcommand{\opussuccessoverall}{94\%\xspace}
\newcommand{\opustimeoverall}{15.3~min\xspace}
\newcommand{\opuscostoverall}{\$13.65\xspace}
\newcommand{\sftimeoverall}{7.2~min\xspace}
\newcommand{\sfcostoverall}{\$5.61\xspace}

\title{\bench: A Benchmark Suite for \\ Rust System Verification}

\author{%
  Chenyuan Yang \\
  University of Illinois Urbana-Champaign \\
  \texttt{cy54@illinois.edu} \\
  \And
  Natalie Neamtu \\
  Carnegie Mellon University \\
  \texttt{nneamtu@andrew.cmu.edu} \\
  \AND
  Chris Hawblitzel \\
  Microsoft Research \\
  \texttt{Chris.Hawblitzel@microsoft.com} \\
  \And
  Jacob R. Lorch \\
  Microsoft Research \\
  \texttt{lorch@microsoft.com} \\
  \AND
  Shan Lu \\
  Microsoft Research \& University of Chicago \\
  \texttt{shanlu@microsoft.com} \\
}

\begin{document}
\maketitle

\begin{abstract}
Large language models (LLMs) have shown impressive capability to understand and develop code. 
However, their capability to rigorously reason about and formally prove the correctness of software systems remains in question. 
Current benchmarks for formal proof generation focus on isolated tasks, typically algorithmic-style problems utilizing basic data structures. 
We curate a new benchmark suite for system-level proof generation, \bench, which consists of \tasks proof tasks extracted from \projects open-source Verus-verified Rust systems, covering operating systems, storage systems, memory allocators, and more.
These tasks are much more complicated than those in previous benchmarks, containing on average more than 20$\times$ the formal specification (in Lines of Code) and requiring a much broader array of verification techniques. 
We evaluate six LLMs on \bench (\gmini, \gfive, \sfour, \sfourfive, \opusfourfive, Qwen3.5), and demonstrate that different models and agentic setups lead to vastly different success rates and costs.
We also show that the best models possess impressive capability in writing correctness proofs and hence should inspire new ways of developing trustworthy software. Meanwhile, some difficult tasks in \bench stress even the best model, forcing long reasoning time (over an hour on the hardest tasks, \opustimeoverall per task on average), high token cost (\opuscostoverall per task on average), and proof failures, which we hope will guide future improvements in models and agents. This is particularly true for a harder no-lemma version of \bench, where the helper lemmas written by human experts are removed: there, the best model proves only 41\% of the tasks from the hardest project, at \$96 per task.
\end{abstract}

\section{Introduction}
\label{sec:intro}
In the past few years, two contrasting software development methodologies have progressed. On the one hand, AI coding agents~\cite{yang2024swe,xia2025live} can quickly produce a lot of code with little human support. However, they do not offer any correctness guarantees, a big concern for reliability-critical software. On the other hand, system verification techniques are getting mature after decades of research. Recent work~\cite{verifiedstorage, vest, atmososp25, verussosp24, cortenmm} demonstrates the feasibility of developing formal correctness specifications and proofs for large-scale system software in a popular programming language (i.e., Rust) using tools such as Verus~\cite{verussosp24,verus-oopsla}.
However, such development requires great effort from formal verification experts. Naturally, one wonders
if the complementary strengths of LLMs and formal verification can together unlock practical, large-scale systems development with machine-checked proofs.

Towards this goal, many recent projects have explored using LLMs for \textit{proof generation}, where LLMs must produce proof annotations to convince the verifier that a piece of code matches its given correctness specification.
Various techniques (e.g., agent systems, model fine-tuning, RAG) have been tried, and yet it is still unclear whether LLMs can generate proofs from scratch for real-world systems: 
some work focused on proof-oriented languages instead of general programming languages~\cite{saikat.icse25, dafnybench}; some focused on small programs like binary search and data-structure implementation~\cite{autoverus, dafnybench, rvbench, safeiclr25, alphaverus, veristruct2025}; some looked at large systems but focused on finishing or repairing a local proof given all the other correct proofs developed by human experts in the system \cite{rvbench, mugnier2025laurel}.

We believe a critical missing piece is a benchmark suite that represents proof tasks in real-world large systems. Without such a benchmark, it is not clear how well LLMs and related techniques work for system verification, not to mention what aspects particularly require further research.

To fill this gap, we curate \bench, a benchmark suite for proof tasks in real-world system projects written in Rust (\S\ref{sec:bench-setup}). 
It consists of \tasks proof tasks extracted from \projects open-source Verus-verified system projects, spanning operating systems, storage systems, distributed systems, and more. 
This includes 157 tasks extracted from a project that was not released online until November 2025 and hence is not in the training data of the models we evaluate.
Every task in \bench corresponds to one proof function or executable Rust function in the original project, with all cross-file dependencies extracted into a stand-alone file that can be individually compiled and verified. 
Each task file contains \textbf{no} proof annotations from the original project (i.e., no proof example for LLMs), and hence brings us closer to testing LLMs' innate system-proof-writing capabilities. Among these \tasks proof tasks, original human developers used lemmas in the proof of \noltasks tasks. 
As a result, \bench offers two levels of difficulty: normal tasks keep the signatures (but not the bodies) of these helper lemmas, while no-lemma tasks remove them, so models need to decompose the proof by themselves. We also release a harder subset, \benchhard, which contains the tasks that none of the evaluated models can prove, plus the no-lemma tasks of the hardest project.

We perform a quantitative comparison (\S\ref{sec:bench-analysis}) between \bench and existing Verus benchmarks (VerusBench \cite{autoverus}, HumanEval-Verus \cite{humanevalverus}, AlphaVerus \cite{alphaverus}).
Unsurprisingly, \bench tasks are much larger, containing 498 lines of specification per task on average and requiring original developers to write 54 lines of proof annotations per task. In comparison, the three existing benchmarks contain only 8 -- 25 lines of specification per task and require only 4 -- 29 lines of proof annotations in their ground truth solutions.
Our analysis also shows that proofs in verified systems rely much less on certain techniques emphasized in prior benchmarks (e.g., loop invariants) and rely much \textit{more} on other proof techniques not exercised by existing benchmarks (e.g., proof by induction, proof by contradiction, special non-linear arithmetic and bitvector solvers).

We have used \bench to evaluate the proof-generation capability of five different models (\gmini, \gfive, \sfour, \sfourfive, \opusfourfive), plus the open-weight Qwen3.5, under two agentic setups (\S\ref{sec:agent-protocol}), one generic coding agent (e.g., GitHub Copilot CLI~\cite{github_copilot_cli}) 
and one planner/action agent we customized for Verus.
Our evaluation (\S\ref{sec:eval}) shows that different model/agent combinations offer vastly different success rates, ranging from a mere 17\% (\gmini with coding agent) to an impressive 94\% (\opusfourfive with coding agent). Intrigued by the high success rate of some of these models, we further evaluated and observed that LLMs completed 33 proofs that had not yet been finished by human experts in an operating system project, further demonstrating the potential of using LLMs to develop formally verified systems. Meanwhile, our evaluation also exposes directions for future research. For example, under the same coding agent, \opusfourfive introduces much larger time cost (more than 4$\times$) and money cost (more than 10$\times$) than less capable models like \gmini and \gfive, inviting future work on balancing capability with cost and speeding up models' formal reasoning. Furthermore,
even the most capable models, \sfourfive and \opusfourfive, struggle at some of the difficult tasks. In the no-lemma version of \bench, where stubs for human-developed helper lemmas are removed, the success rates of these two models drop from 71\% and 92\% down to 57\% and 87\%, respectively, on the same \noltasks tasks. On the hardest project, Anvil Controller, \opusfourfive proves only 41\% of these no-lemma tasks, spending \$96 per task on average. Future research is needed to help these models on topics like proof/specification abstraction, inductive invariants, etc. We have released \bench, our agent, and all model trajectories at \benchurl.
\section{\bench Setup and Study\label{sec:background}}

\MyPara{Background} Verus~\cite{verus-oopsla,verussosp24} is a tool designed to offer high-performance verification for system software written in Rust. 
Using Verus, developers can provide a \emph{specification} for any functions they choose to verify in the form of function pre-conditions (i.e., what properties the function arguments should satisfy) and post-conditions (i.e., what properties the return value should satisfy) in Rust syntax. 
Developers can then provide \emph{proof annotations}, also in Rust syntax, to aid the Verus verifier in proving those obligations. An example of Verus specification and proof is shown in Appendix \ref{sec:example}.

\subsection{\bench Construction}
\label{sec:bench-setup}

Our goal is to choose large-scale Verus-verified systems and turn each function with proof annotations, including Rust executable functions and Verus proof functions (i.e., lemmas), into a stand-alone task.

\MyPara{Project selection.}
We select the \projects open-source Verus-verified
system projects shown in Table~\ref{tab:systemlist}.
Among these \projects projects, five are the projects presented in the Verus paper \cite{verussosp24}. In addition, Vest~\cite{vest} is a more recent project about verified parsers and serializers; Atmosphere~\cite{atmososp25} had {\bf not} yet been released to the public when we received the code from its authors and started our experiments in late November 2025 and hence is not in the training dataset of any models we will evaluate (it has since been released under the MIT license~\cite{atmosgithub}, and its authors gave us written permission to redistribute the task files extracted from it); Anvil \cite{anvil} proves not only safety properties but also liveness properties. We separate Anvil proof tasks into two separate groups due to their completely different code and specification properties: Anvil temporal-logic library and Anvil Kubernetes controller (AL and AC respectively in this paper). 
Together, these projects implement a wide range of systems, are developed by many different authors, and have all been featured in recent systems research papers \cite{verussosp24, vest, anvil, verifiedstorage, atmososp25}. All eight projects use the MIT license (NRKernel also allows Apache-2.0), and each released task file keeps the original license and copyright notice (Appendix~\ref{app:license}).

\MyPara{Benchmark extraction.} Once a Verus-verified system project is selected, our benchmark extraction goes through the following steps.
First, we manually filter out {\it trivial} functions that do not require
proof annotations. We also remove functions and modules that the original developers suggested we skip.
We then extract each non-trivial function \verb|F|, together with all its cross-file and cross-module code dependencies, into a stand-alone verified Rust file \verb|F_verified.rs|. To accomplish this, we run Verus with \verb|log-all| mode, which enables Verus to log the abstract syntax trees (ASTs) of all the code structures that Verus uses to prove a target function. We then process that log file to produce a stand-alone Verus-verified Rust file. 
We intentionally put each task into a single file: collecting the code a function depends on is generic code navigation that can be automated (as \codeIn{log-all} does), so \bench focuses on writing the proof. Since every definition needed by the proof is kept, a failure on \bench is a failure of proof writing, not of finding code.

\MyPara{Normal tasks and No-lemma tasks.} We remove all the proof annotations in the stand-alone file \verb|F_verified.rs|, so that we can test LLMs' true capability in writing proof annotations without hints (i.e., proof written by human experts in the same project). We use Verus annotations to allow the verifier to skip verification and assume the specification for all functions in the file except the target proof-task function \verb|F|. This produces our normal task file \verb|F_unverified.rs|.

In about half of these task functions (\noltasks out of \tasks), one or more lemmas were used by the original human experts to prove the correctness of function \verb|F|. In the normal task file, we leave all these lemmas for LLMs to use -- the proof bodies of these lemmas are removed in order to not provide extra hints. For each such function, we also create a harder \textit{no-lemma} task \verb|F_nolemma_unverified.rs| in which all lemmas are removed. Thus, the model must produce the full proof for \verb|F| from scratch.

\MyPara{Hard subset and repo-level track.} \benchhard contains the normal tasks that none of our model/agent configurations could prove, plus all no-lemma tasks from Anvil Controller (AC), the hardest project (Appendix~\ref{app:stats}). The repo-level track puts each task back into its original repository, so agents also need to find the code dependencies themselves (pilot in Appendix~\ref{sec:eval-detailed-investigations}).

\subsection{\bench Characteristics and Comparison}
\label{sec:bench-analysis}

\begin{table}[t]
    \caption{Open-source Verus-verified Rust systems in our study, with average line-of-code statistics for \bench tasks per project. ``Lemma'' is the LoC of helper-lemma signatures retained in the benchmark (their bodies are stripped); ``Target'' is the proof annotations on the task's target function. ``Other'' is the residual (executable code, attributes, blank lines). ``License'' is the license of the original project (Appendix~\ref{app:license}).}
    \label{tab:systemlist}
    \label{tab:stats}
    \tablestyle
    \setlength{\tabcolsep}{3pt}
    \begin{tabular}{lllrrrrrrl}
    \toprule
     & & & & & \multicolumn{2}{c}{Proof} & & & \\
    \cmidrule(lr){6-7}
    System & Abbr. & Description & \#Tasks & Spec & Lemma & Target & Other & Total & License \\
    \midrule
    Anvil Library    & AL & Temporal-logic library      & 104 & 29   & 10 & 9  & 36   & 84   & MIT \\
    Anvil Controller & AC & A Kubernetes controller     & 63  & 2037 & 33 & 75 & 1871 & 4016 & MIT \\
    IronKV           & IR & Sharded key-value store     & 118 & 141  & 8  & 40 & 225  & 414  & MIT \\
    Memory Allocator & MA & Mimalloc in Rust            & 89  & 51   & 4  & 49 & 34   & 138  & MIT \\
    Node Replication & NO & Replication library         & 29  & 41   & 2  & 25 & 32   & 99   & MIT \\
    NRKernel         & NR & OS page table               & 204 & 676  & 13 & 44 & 458  & 1190 & MIT/Apache-2.0 \\
    Atmosphere       & OS & Operating system            & 157 & 731  & 13 & 60 & 499  & 1302 & MIT \\
    Storage          & ST & Persistent-memory storage   & 63  & 249  & 9  & 30 & 258  & 546  & MIT \\
    Vest             & VE & Binary parser \& serializer & 22  & 29   & 5  & 14 & 51   & 99   & MIT \\
    \midrule
    \multicolumn{4}{l}{Avg.} & 498 & 11 & 43 & 402 & 954 & \\
    \bottomrule
    \end{tabular}
\end{table}

The \verb|F_verified.rs| files in \bench offer a good opportunity to understand what it takes for human experts to prove a function in a system project and to compare the characteristics with other Verus benchmarks.
We compare \bench with three prior Verus proof-synthesis benchmarks: VerusBench~\cite{autoverus}, HumanEval-Verus \cite{humanevalverus} and AlphaVerus \cite{alphaverus}. The former two mainly converted subsets of non-verification benchmarks like MBPP \cite{DBLP:journals/corr/abs-2108-07732} and HumanEval \cite{humaneval} into Rust programs, and manually provided Verus specification and proof. AlphaVerus is translated from another verification benchmark suite, DafnyBench \cite{dafnybench}.

\begin{table}[t]
    \caption{Per-task statistics, averaged across all benchmark tasks in four Verus proof-synthesis benchmarks.}
    \label{tab:prior-benchmark-comparison}
    \tablestyle
    \begin{tabular}{lrrrrrrrr}
    \toprule
        &
        & \multicolumn{2}{c}{Avg. LoC per task}
        & \multicolumn{2}{c}{Avg. proof LoC}
        & \multicolumn{2}{c}{Avg. count per task}
        & \\
    \cmidrule(lr){3-4} \cmidrule(lr){5-6} \cmidrule(lr){7-8}
    Benchmark & \#Tasks & Spec & Proof & Loop-inv & Non-loop & Loops & Helper lemmas & Total LoC \\
    \midrule
    VerusBench~\cite{autoverus}                      & 150     & 8   & 11 & 7 & 4  & 1.6  & 0.07 & 52  \\
    HumanEval-Verus~\cite{humanevalverus}            & 74      & 25  & 29 & 7 & 22 & 1.3  & 0.96 & 166 \\
    AlphaVerus translations~\cite{alphaverus}        & 247     & 8   & 4  & 3 & 1  & 0.9  & 0.04 & 38  \\
    \midrule
    \textbf{\bench}                                  & \tasks  & 498 & 54 & 1 & 53 & 0.08 & 2.6  & 954 \\
    \bottomrule
    \end{tabular}
\end{table}

\MyPara{Size of Specifications} As shown in Table~\ref{tab:prior-benchmark-comparison}, \bench tasks contain many more lines of specification (498 on average) than the existing Verus benchmarks, which average only 8 -- 25 lines of specification.
This difference is expected. Prior benchmarks mostly encode one self-contained programming problem or one data-structure implementation at a time. In contrast, each function in a large system often involves many different data structures and the system property can be complicated to express. For example, the specification of many tasks in Anvil Controller (AC) involves an entire system state machine with transition properties. Consequently, each AC task averages 2037 lines of specification (Table \ref{tab:systemlist}).
Reading and understanding the entire specification is a challenge for proof development in system projects!

\MyPara{Proof Size and Number of Lemmas} As also shown in Table \ref{tab:prior-benchmark-comparison}, many more lines of proof annotations are needed to prove a task in \bench (54 lines) than other Verus benchmarks (4--29 lines) based on human-written ground-truth proof.

When we dive deeper, we see much fewer loop invariants and many more helper lemmas in each \bench task. System projects contain relatively few loops and hence only 1 line of loop-invariant is contained per task on average in \bench (less than 2\% of total proof). In contrast, much previous work focused on loop invariant generation \cite{autoverus, DBLP:conf/cav/ChakrabortyGU20} and the proofs in previous benchmarks contain much higher portion of loop invariants, from 24\% in HumanEval-Verus to 64\% in VerusBench.
Another big difference is in the average number of helper lemmas used by human experts to prove each task: 2.6 vs. 0.04--0.96 in other benchmarks. This is because, in large systems, human experts often decompose a complicated proof into smaller pieces, each being a helper lemma.

\begin{figure}[t]
    \centering
    \includegraphics[width=\linewidth]{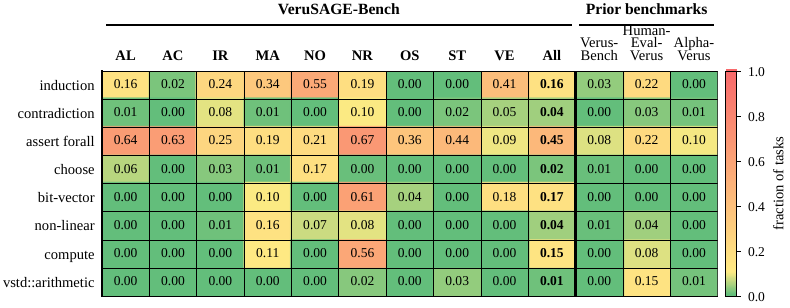}
    \caption{Fraction of task files in each benchmark that use selected Verus proof features.}
    \label{fig:prior-bench-heatmap}
\end{figure}

\MyPara{Proof-feature coverage.}
Figure~\ref{fig:prior-bench-heatmap} shows the fraction of task files in
each benchmark that touch eight important Verus proof features (induction,
proof-by-contradiction, \codeIn{assert forall}, \codeIn{choose}, the
bit-vector and non-linear-arithmetic solvers, \codeIn{by(compute)}, and
\codeIn{vstd::arithmetic}). \bench exercises every feature at non-trivial
frequency: \codeIn{assert forall} appears in nearly half of its tasks,
the bit-vector solver in 17\%, and \codeIn{by(compute)} in 15\%. The
small-program benchmarks barely touch the specialized solvers:
bit-vector is absent across all three, \codeIn{choose} appears in only
one VerusBench task, and \codeIn{by(compute)} appears only in
HumanEval-Verus (8\%). HumanEval-Verus has some recursive proof helpers
(induction in 22\% of files, mostly base-case-style); AlphaVerus
rarely uses induction.
\section{Evaluation Protocol and Agent Settings}
\label{sec:agent-protocol}

We use \bench to evaluate five LLMs in two agent settings: Hands-Off and Hands-On.

\MyPara{Models.}
We evaluate five state-of-the-art LLMs. From OpenAI, we consider \gmini (April 2025)~\cite{o4mini} and \gfive (August 2025)~\cite{gpt5}: \gmini, from the o-series, is designed to think longer than GPT-4/4o, while \gfive aims for further gains in reasoning capability. From Anthropic, we include three Claude models, \sfour (May 2025)~\cite{sonnet4}, \sfourfive (September 2025)~\cite{sonnet45}, and the more recent \opusfourfive (November 2025)~\cite{opus45}; at release, Anthropic described \sfourfive as the ``best coding model in the world''~\cite{sonnet45}, and \opusfourfive as their flagship for top-tier reasoning and coding.
To see how open models perform, we also evaluate Qwen3.5 (397B-A17B)~\cite{qwen35}, an open-weight model, under the same setting; its results are in Appendix~\ref{app:qwen}.

\MyPara{Hands-Off.} The model runs under a generic coding agent (GitHub Copilot CLI~\cite{github_copilot_cli} for \gfive/\sfour/\sfourfive/\opusfourfive; OpenAI Codex CLI~\cite{openai_codex} for \gmini) with a short prompt that (i) tells it not to cheat, and (ii) gives it access to Verus, a Verus cheat checker (which ensures agents cannot modify the task), and a local copy of the Verus standard library (\codeIn{vstd}). Every run happens in a container with no proof from the target project present. The coding agent decides when to end a task.

\MyPara{Hands-On.} A planner-dispatcher agent invokes 16+ specialized \emph{action agents} (logical reasoning, arithmetic solvers, proof-context manipulation, quantifier instantiation, etc.), maintains a history of past proof attempts, and uses a candidate-selection policy that generalizes AutoVerus's error-count heuristic~\cite{autoverus}. Each task runs until Verus successfully verifies the proof without cheating, or up to 20 proof steps or 20 minutes, whichever comes first. \tool adds roughly 15{,}600 lines of Python on top of AutoVerus. Full implementation details are in Appendix~\ref{app:agent}. Each configuration runs once per task; we report confidence intervals and significance tests in Appendix~\ref{app:stats}.
\section{Experimental Results \label{sec:evaluation}}
\label{sec:eval}

We use \bench to answer: (i) how often do current LLMs succeed, and how does that differ across models and agents (\S\ref{sec:eval-success-rate})? (ii) when and how do they succeed (\S\ref{sec:eval-when-succeed})? (iii) when and why do they fail (\S\ref{sec:eval-when-fail})? (iv) what is the cost (\S\ref{sec:eval-cost})? Detailed result breakdown among the 9 projects in \bench, alternative agent settings, ablations, and additional case studies, including LLMs successfully finishing proofs human experts have not yet finished, are deferred to Appendix~\ref{app:extended-eval}.

\subsection{How often do LLMs succeed on \bench?}
\label{sec:eval-success-rate}

\begin{table}[t]
    \caption{Overall evaluation results on \bench (Off/On: Hands-Off/Hands-On agent). The ``95\% CI'' row is the half-width of the 95\% confidence interval of the success rate, computed over tasks (Appendix~\ref{app:stats}).}
    \label{tab:allresults}
    \tablestyle
    \begin{tabular}{lrrrrrrrrrrr}
    \toprule
    &\multicolumn{9}{c}{Normal tasks (\tasks)} & \multicolumn{2}{c}{No-lemma tasks (\noltasks)}\\
    \cmidrule(lr){2-10} \cmidrule(lr){11-12}
    &\multicolumn{2}{c}{\gmini}
    &\multicolumn{2}{c}{\gfive}
    &\multicolumn{2}{c}{\sfour}
    &\multicolumn{2}{c}{\sfourfive}
    &\multicolumn{1}{c}{\opusfourfive}&\multicolumn{1}{c}{\sfourfive}&\multicolumn{1}{c}{\opusfourfive}\\
    \cmidrule(lr){2-3} \cmidrule(lr){4-5} \cmidrule(lr){6-7} \cmidrule(lr){8-9} \cmidrule(lr){10-10} \cmidrule(lr){11-11} \cmidrule(lr){12-12}
    &Off&On    &Off&On  &Off&On     &Off&On     &Off & Off & Off\\
    \midrule
Success rate    &17\% & 41\% & 50\% & 56\% & 64\% & 58\% & 81\% & 67\% & 94\% & 57\% & 87\% \\
95\% CI ($\pm$pp) & 2.5 & 3.3 & 3.4 & 3.3 & 3.2 & 3.3 & 2.6 & 3.2 & 1.6 & 4.5 & 3.1 \\
Avg.\ time (min) & 3.3 & 12.8 & 3.7 & 12.7 & 9.2 & 7.1 & 7.2 & 6.9 & 15.3 & 10.9 & 30.8 \\
Avg.\ cost (\$) & 0.8 & 0.7  & 1.3 & 0.5  & 6.7 & 1.7 & 5.6 & 1.5 & 13.7 & 10.8 & 52.0 \\
    \bottomrule
    \end{tabular}
\end{table}

Table~\ref{tab:allresults} summarizes the success rate of each model under the Hands-Off and Hands-On agent settings. Full result breakdowns across every project inside \bench are in Tables~\ref{tab:details-cli}, \ref{tab:details-verusys}, and \ref{tab:details-opus} in Appendix~\ref{app:extended-eval}. This result shows several interesting trends.

First, system proof tasks are indeed challenging, particularly for smaller or older models like \gmini. Previous research has shown that through careful agent system building (e.g., the AutoVerus system \cite{autoverus}) and fine-tuning \cite{safeiclr25, alphaverus}, even older models like GPT-4o \cite{autoverus} and fine-tuned models based on DeepSeekCoder-33B \cite{safeiclr25} and Llama-3.1-70B \cite{alphaverus} can achieve over 70\% success rate on previous benchmarks like VerusBench. For comparison, we applied \gmini with the AutoVerus agent system on our benchmarks, and the success rate is merely 20\%. This difference highlights the importance of benchmark building, and calls for more research on using smaller models to write system proofs.

Second, frontier models with state-of-the-art coding agents are capable of writing system proofs!
The most surprising result to us is that strong Anthropic models with the Hands-Off setup prove a large majority of \bench tasks without any human intervention. \sfourfive solves 81\% of the \tasks tasks and the more recent \opusfourfive pushes this to \opussuccessoverall (797/849). 
These wins are \textbf{not} explained by memorization: the LLM proof differs greatly from the human proof (\S\ref{sec:eval-when-succeed}), and Atmosphere (OS) — at 83\% for \sfourfive and 93\% for \opusfourfive — was not public before either model's training cutoff (we received the code from its authors in late November 2025), yet both models do as well on it as on the whole benchmark. These differences between models are also statistically significant: neighboring models differ by 13--33 percentage points, several times the width of their confidence intervals. These results indicate that it is feasible to leverage LLMs to write formally verified software, and that formal methods can indeed serve as a guardrail for LLMs' code synthesis. Of course, the cost of these frontier models is still a concern (\S\ref{sec:eval-cost}).

Of course, the No-Lemma results in Table \ref{tab:allresults} show that if human experts do not help decompose complicated proof tasks into smaller pieces (i.e., lemmas), even the best models today suffer either drastically decreased success rates or substantially increased time and money costs. Specifically, among the \noltasks tasks whose original proof uses helper lemmas, \sfourfive's success rate drops from 71\% to 57\%. \opusfourfive's success rate drop is not as big, from 92\% to 87\%. However, its time and money cost increase dramatically from \opustimeoverall and \opuscostoverall to roughly 31 minutes and \$52 per task on average ($\sim$2$\times$ slower, $\sim$4$\times$ more expensive). The drop is the largest in Anvil Controller (AC), whose human proofs rely on long chains of lemmas: there, \opusfourfive proves only 41\% of the no-lemma tasks at \$96 per task, and \sfourfive only 19\% (Table~\ref{tab:nolemma}). Furthermore, no configuration can prove 37 of the normal tasks. Together with the 58 no-lemma AC tasks, they form \benchhard, where the best configuration proves 25\% of 95 tasks (Appendix~\ref{app:stats}).

Third, there is a lot of room for improving customized agent systems. As shown in Table~\ref{tab:allresults}, the customized agent system (Hands-On) helps less-capable models like \gmini and \gfive, boosting their success rate from 17\% and 50\% to 41\% and 56\%, respectively. However, this boosting effect disappears for more-capable models like \sfour and \sfourfive, to the point that we decided to skip running \opusfourfive in Hands-On mode to save cost. Instead, for these more capable models, they seem to benefit more from generic coding agents that offer them the freedom to explore the code base of the proof task and Verus standard library repeatedly, and to keep updating proof and checking the proof by running Verus at the model's own pace. How to build customized agents to better support these frontier models and further improve their success rates remains a question.
Overall, the model matters more than the agent: with the same Hands-Off agent, success rates range from 17\% to 94\% across models, while changing the agent moves a model by at most 24 points. The open-weight Qwen3.5 lies between \gmini and \gfive (25\% Hands-Off, 45\% Hands-On; Appendix~\ref{app:qwen}).

Note that the Hands-Off coding agent still has Verus-related tool support. Even \sfourfive rarely generates correct proof at the first attempt, which occurred for only 3\% of the tasks. In most cases, the model needs to repeatedly run Verus to see what the problems with its current proof are and adjust accordingly. When \opusfourfive worked on No-Lemma tasks, its trajectory recorded more than 16 Verus runs per task on average. The Verus standard library and the cheat checker also helped.

Finally, the difficulty of proof tasks coming from different open-source projects differs a lot. As we present in Tables~\ref{tab:details-cli}, \ref{tab:details-verusys}, and \ref{tab:details-opus} in Appendix~\ref{app:extended-eval}, some projects like Node Replication (NO) and Vest (VE) are relatively easy for all models and agents; meanwhile, the hardest project across all models and agents is Anvil Controller (AC). \sfourfive and \opusfourfive both achieve 100\% success rate among tasks from the same three projects, Anvil Library, Node Replication, and Vest.

\subsection{When do LLMs succeed and how?}
\label{sec:eval-when-succeed}

\begin{figure}[t]
    \centering
    \begin{minipage}[b]{0.43\linewidth}
    \centering
    \includegraphics[width=\linewidth]{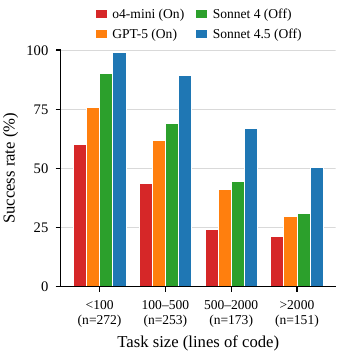}
    \caption{Success rate by task size (each model under its best agent setting).}
    \label{fig:sizesuccess}
    \end{minipage}\hfill
    \begin{minipage}[b]{0.54\linewidth}
\begin{lstlisting}[language=Verus,style=paperlisting,escapechar=\%]
assert forall|s: Seq<A>, v: A|
  s.len() > 0 && s[0] != v implies
  s.skip(1).contains(v) == s.contains(v) by {
  broadcast use vstd::seq_lib::lemma_seq_skip_contains
%\Redundant%  if s.skip(1).contains(v) {
%\Redundant%    assert(exists|i: int| 1 <= i < s.len()
%\Redundant%                && s[i] == v);
%\Redundant%    assert(s.contains(v));
%\Redundant%  }
%\Redundant%  if s.contains(v) {
%\Redundant%    let i = choose|i: int| 0 <= i < s.len()
%\Redundant%                && s[i] == v;
%\Redundant%    if i == 0 {
%\Redundant%      assert(s[0] == v);
%\Redundant%      assert(false); // a contradiction
%\Redundant%    } else {
%\Redundant%      assert(1 <= i < s.len() && s[i] == v);
%\Redundant%      assert(s.skip(1).contains(v));
%\Redundant%    }
%\Redundant%  }
}
\end{lstlisting}
    \vspace{-4pt}
    \caption{In this \sfourfive proof for a lemma in Atmosphere (OS), the {\setlength{\fboxsep}{1pt}\colorbox{red!10}{shaded}} lines are unnecessary: the proof still verifies without Lines 5--20, since Verus derives the goal on Lines 1--3 directly from the lemma on Line 4.}
    \label{fig:chatty}
    \end{minipage}
\end{figure}

\MyPara{What correlates with success?}
We tested whether success correlates with task size (LoC of \verb|F_unverified.rs|). Using the Point-Biserial Correlation Coefficient~\cite{enwiki:1289424476}, all four models show a negative correlation in the range $-0.27$ to $-0.50$ (P-values all $\le$ $4.7\mathrm{e}{-16}$, highly significant): smaller tasks are easier. Figure~\ref{fig:sizesuccess} visualizes the success rate per size bucket. Note that we consider only normal tasks (i.e., not no-lemma setting) in this study, and we did not include \opusfourfive as it succeeds in most normal tasks.

\MyPara{How do LLM and human proofs differ?}
We did a deep dive on the proofs of Hands-Off \sfourfive.

\underline{Proof features.} Human and LLM proofs use similar but not identical strategies. The LLM uses proof-by-contradiction more, the non-linear prover less, and \codeIn{vstd::arithmetic} more. In MA, among the 11 tasks where developers use the non-linear prover and \sfourfive does not, \sfourfive fails 4 and leverages \codeIn{vstd::arithmetic} on the rest — suggesting a weakness in non-linear arithmetic, compensated by strong search for relevant helper lemmas.

\underline{Proof overlap.} The proof lines that \sfourfive adds have a median overlap (Jaccard similarity, over 830 tasks) of only 6.5\% with those added by human experts, and 86\% of its proofs share less than a quarter of their lines with the human proof.

\underline{Proof length.} Across the \sfourfivetasks tasks where \sfourfive (Hands-Off) succeeds, humans add a median (mean) of 9 (17.3) lines of proof; \sfourfive adds 24 (44.2). Very rarely is the human proof longer.

This longer proof length reveals a weakness in \sfourfive (which \opusfourfive suffers from as well): \sfourfive actually does not know for sure what properties can or cannot be verified by Verus and its underlying SMT prover unless proof hints are provided. In the example shown in Figure \ref{fig:chatty}, the properties asserted on Lines 8, 15, and 18 can easily be derived by the verifier from the properties asserted on Lines 6, 14, and 17, respectively. In fact, all of Lines 5--20 are unnecessary, as the verifier is able to derive the property to prove (Lines 1--3) from the \texttt{seq\_lib} lemma used on Line 4. In many cases, models simply guess what properties are needed to finish the proof chain and leave it to Verus to decide whether the proof succeeds or not. 

Chatty proofs waste tokens (i.e., cost) and also hurt proof readability. The lack of understanding about when proof annotations are needed could also affect models' success rate for complicated tasks. This is a topic worth investigating in future research.

\vspace{-1em}%

\subsection{When and Why do LLMs fail?}
\label{sec:eval-when-fail}

\MyPara{Why does \sfourfive fail?} We carefully checked all the normal tasks that \sfourfive failed to prove and identified three coding and verification patterns that remain challenging even for capable models.

\emph{Inductive invariants (AC).} Human proofs in Anvil Controller (AC) often establish an invariant by strengthening it until it is inductive. This could be counter-intuitive to models or novice human provers: why try to prove a stronger invariant than the desired one? Automatically inferring an inductive invariant is an open research area~\cite{basilisk,i4,distai,duoai}, and remains out of \sfourfive's reach. This challenge is also the reason for \opusfourfive to fail 17\% of tasks in AC --- its worst project.

\emph{Procedural macros (ST).} Verified storage (ST) uses a procedural macro to synthesize size/alignment functions via \verb|#[repr(C)]|. \sfourfive, though given access to the macro definition, cannot expand it (or did not realize the need to expand it) to learn the resulting definitions.

\emph{Deep abstraction (NR).} NRKernel uses many \verb|closed| spec functions, which are heavily featured in Atmosphere (OS) as well; proofs in other modules must rely on lemmas about those hidden definitions. \sfourfive tends to insist on seeing the hidden bodies and misses that the lemmas suffice.

\MyPara{Why does \gmini fail so often?} \gmini struggles with Rust and Verus syntax and hallucinates lemmas. In Hands-On mode, its proof candidates contain syntax errors 7.8 times per task on average; common errors include ``unexpected token'' (21\% of tasks) and ``cannot find value/function'' (33\%). In contrast, only 10\% of \sfourfive tasks produced any such error.

\subsection{How much time and money do LLMs spend?}
\label{sec:eval-cost}

\begin{figure}[t]
    \centering
    \includegraphics[width=\linewidth]{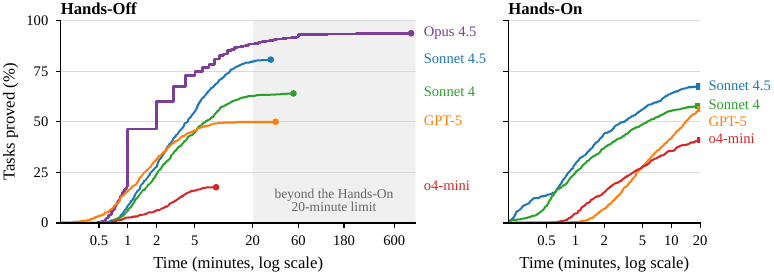}
    \caption{\% of tasks accomplished within a certain time (min), for Hands-Off (left) and Hands-On (right), on the same log scale. Hands-On stops at its 20-minute limit; the shaded region in the left panel marks times beyond it. Markers show the final success rate of each model; times of \opusfourfive are recorded per minute.}
    \label{fig:cdf-time}
\end{figure}

\MyPara{Money.} Table \ref{tab:allresults} shows that Hands-On costs less than Hands-Off in general because, compared with a generic coding agent, the customized agent uses focused prompts and strict requirements on what to do in each step, leaving little room for the models to read various information and freely iterate. 

Across different models working with a generic coding agent, the cost across \gmini, \gfive, and the two Sonnet models roughly reflects their token rate differences. The cost per 1 million input tokens is \$1.1 for \gmini, \$1.25 for \gfive, and \$3 for the two Sonnet models. 

An outlier, and somewhat of a concern, is \opusfourfive. Although its input token price is 1.67$\times$ that of \sfourfive (\$5 vs. \$3), its average per-task cost is 2.4$\times$
that of \sfourfive. This difference under the No-Lemma setting further increases to 4.8$\times$: \$52.0 for \opusfourfive vs. \$10.8 for \sfourfive. In one task under AC, \opusfourfive failed to prove the main property and simply explained that ``SMT solver is not able to connect the facts'' after spending more than \$500. 
Even with the high success rate, the huge cost of \opusfourfive is probably not affordable for many developers. 
Cheaper options exist: Hands-On \gfive proves 56\% at \$0.52 per task, and Hands-Off Qwen3.5 costs about \$0.7 per task (Appendix~\ref{app:qwen}), though both succeed less often.

\MyPara{Time.} Considering only successes, \sfourfive is the fastest: Hands-On \sfourfive proves 40\% of tasks in under 2 minutes; Hands-Off \sfourfive proves 60\% in under 6 minutes (Figure~\ref{fig:cdf-time}). \opusfourfive trades speed for success: it averages \opustimeoverall per task, twice \sfourfive's \sftimeoverall, but converts that extra wall-clock into 13 percentage points of additional success.

Speed of giving up differs by setting and model. Hands-On enforces $\ge 20$ steps or $\ge 20$ minutes before giving up. Hands-Off leaves the decision to the model; \gmini and \gfive give up much more readily than Sonnet. On AC, \sfourfive (Hands-Off) runs Verus 13.6 times per task averaging 12.5 minutes, while \gmini runs it 2.6 times and finishes in under 3 minutes.
\section{Limitations of \bench}
\label{sec:limit}

\bench evaluates proof synthesis given a fixed, human-authored specification and project structure. There are still several aspects of system verification that \bench does not cover:
\begin{itemize}[itemsep=0pt, parsep=0pt, leftmargin=*]
  \item \textbf{Specification writing.} Whether LLMs can draft pre/post-conditions or whole-system invariants is out of scope; a verifier can check that code meets a spec, but not that a spec matches user intent.
  \item \textbf{End-to-end project lifecycle.} It is hard to tell how a human prover would use and maintain proofs generated by LLMs, particularly when those proofs have not yet fully passed Verus.
  \item \textbf{Multi-file proof composition.} We extracted cross-file dependencies into every stand-alone task file. \S\ref{sec:eval-detailed-investigations} shows that token/cost may explode in the whole-project setting for large codebases.
  \item \textbf{Advanced Verus features.} Tokenized state-machine macros and permissioned APIs for unsafe datatypes are excluded from \bench (affecting MA and NO). These remain open targets.
  \item \textbf{Closed-source or post-training-cutoff projects.} Atmosphere is our one such project. As models keep updating, more ``out-of-training'' projects are needed to separate capability from memorization. Atmosphere has since become public, but our extraction pipeline works for any project that builds with Verus, so new projects can be added the same way.
\item \textbf{Repeated runs.} We run every configuration once per task, and measured run-to-run variance only on subsets of tasks (Appendix~\ref{sec:eval-detailed-investigations}).
\item \textbf{No-lemma tasks.} It is not as straightforward to create no-lemma tasks, as occasionally a helper lemma may contain axioms that cannot otherwise be verified. We tried our best in keeping those axiom lemmas (otherwise, the task is impossible to prove), but there could be human mistakes.
\end{itemize}
\section{Related Work}

Since we already discussed and compared other Verus benchmarks in Section \ref{sec:bench-analysis}, we briefly discuss work in the domain of LLM-for-code-proof here. 
\llm{s} have been revolutionizing software engineering~\cite{xia2023keep,tian2024debugbench,yang2023white,yang2025kernelgpt,knighter}, sparking interest in automating formal proof synthesis~\cite{DBLP:journals/corr/abs-2404-09939,autoverus,rvbench,mugnier2025laurel}.
Early work focused on prompting or simple RAG for proof-oriented languages like Dafny~\cite{DBLP:conf/saiv/SunSPB14,DBLP:journals/pacmse/MisuLM024,dafnybench} and F$^*$~\cite{saikat.icse25}.
Recent work like Rango~\cite{Rango} for Rocq improves RAG by dynamically retrieving relevant proofs to guide the solver.
For Verus, \autoverus~\cite{autoverus} establishes a baseline workflow using few-shot prompting and error feedback. Most recently, RagVerus~\cite{rvbench} extends \autoverus by using RAG to fetch dependencies and similar proof examples from the target project repository. RagVerus evaluates its technique on four Verus-verified projects (Verismo~\cite{DBLP:conf/osdi/ZhouACGHC24}, IronKV~\cite{verussosp24}, Vest~\cite{vest}, and a small part of Anvil~\cite{anvil}) with the original multi-file structure unchanged, referred to as RvBench. The methodology of \bench differs from RvBench in that we leverage Verus' built-in log support to extract all the code dependencies out to establish single-file, individually verifiable task files without offering any hints to LLMs (i.e., proof annotations written by human experts in other parts of the project). \bench is also based on a more diverse set of 8 open-source systems.
RagVerus's low success rate ($<20\%$) on system projects indicates that RAG alone is insufficient to enable \autoverus and GPT-4o to tackle system-level proof. With \bench, we were able to conduct much more in-depth analysis of a wide range of models and agents.

Model fine-tuning has also been thoroughly explored. SAFE~\cite{safeiclr25} employs self-evolution to synthesize training data and fine-tune models for Verus proof generation.
AlphaVerus~\cite{alphaverus} bootstraps verified code generation by translating from Dafny and refining candidates via tree search.
Both techniques have small programs in their training set, mainly focus on fine-tuning small models (e.g., DeepSeekCoder-33B), and are evaluated on small programs (VerusBench~\cite{autoverus}, HumanEval-Verus~\cite{humanevalverus}), originated from single-file coding benchmarks \cite{DBLP:journals/corr/abs-2108-07732, humaneval}.
Our work provides benchmark resources and various lessons for these approaches to scale their future efforts to system-level verification.

\MyPara{Lean-style theorem proving.} Many benchmarks target interactive theorem provers like Lean~4~\cite{DBLP:conf/cade/Moura021}, such as miniF2F~\cite{minif2f}, ProofNet~\cite{proofnet}, PutnamBench~\cite{putnambench}, LeanDojo~\cite{DBLP:conf/nips/YangSGCSYGPA23}, and FormalMATH~\cite{formalmath}, where each task is a self-contained mathematical statement, tackled by tactic-search provers such as ReProver~\cite{DBLP:conf/nips/YangSGCSYGPA23}, DeepSeek-Prover~\cite{deepseekprover}, AlphaProof~\cite{alphaproof}, and COPRA~\cite{copra}. \bench differs in three ways. First, Verus discharges proof obligations with an SMT solver, so the key skill is choosing the lemmas and annotations that guide the solver; its standard library has only hundreds of lemmas (vs.\ hundreds of thousands in Mathlib~\cite{mathlib}), and solver failures take extra effort to diagnose, unlike Lean's immediate type-checker feedback. Second, each \bench task comes with about 500 lines of system specification; the closest Lean benchmarks, miniCTX~\cite{minictx} (long context) and FVAPPS~\cite{fvapps} and miniCodeProps~\cite{minicodeprops} (program verification), use small programs written for the benchmark. Third, our tasks come from real systems and involve Rust semantics such as machine arithmetic and bit-level invariants. We see the two lines of work as complementary.
\raggedbottom
\section{Discussion and Conclusion}
\label{sec:discussion}

There are many lessons from building and experimenting with \bench.
The difference in size and feature coverage between \bench and existing benchmarks, as well as the different success rates that models have on them, reflects the importance of benchmark curation.

The high success rates of \opusfourfive and \sfourfive
on \bench are an exciting result for both verification experts and general developers. We believe proof writing is an ideal fit for unreliable LLMs because, once the LLMs start to write not only code but also proofs, the verifier can act as an oracle for determining whether AI-generated code can be trusted. 

Our study also reveals the limitations of the state-of-the-art LLMs and agent systems in proof development, both in terms of proof capability, such as how to devise inductive invariants, how to handle the abstraction in large systems, how to write concise proofs, and how to lower the cost of using models in terms of time and money. Note that the cost problem is not a small one. Spending tens or even hundreds of dollars verifying one function (and doing it every time the function is edited) is unlikely for practical adoption. How to train/fine-tune a low cost model with competitive proof capability; how to design better agent systems and proof-development methodology to further improve the proof-generation capability of models, and to use different models smartly to balance capability and cost; and how to design a better interface between models and human developers so that they can work together for efficient and economic proof development are all interesting and exciting future research directions. We believe \bench will help future research in motivating and evaluating different design tradeoffs.

As mentioned in Section \ref{sec:limit}, there are more challenges in system verification than what was studied in this paper. 
Although models like \opusfourfive can prove most individual, well-specified functions in our setting, we have no evidence that they are good at abstracting and breaking down the project-level verification goal into the right specification at the level of every executable function, and into the appropriate set of manageable proof functions, while fixing code bugs at the same time. These again are directions for future research.

\begin{ack}
This material is based on work supported by the National Science Foundation CISE Graduate Fellowships under Grant No.~2313998. Any opinions, findings, and conclusions or recommendations expressed in this material are those of the author(s) and do not necessarily reflect the views of the National Science Foundation. Shan Lu's research is partially funded by NSF Grant CNS-2313190.
\end{ack}

\bibliographystyle{plainnat}
\bibliography{paper}

\appendix
\section{An Example of Verus Verified Function}
\label{sec:example}

Figure \ref{fig:example} demonstrates a Verus verified function simplified from the NRKernel project. The \texttt{MAX\_PHYADDR} function takes physical address' bit-width as an input and returns the largest possible physical address value. The specification of this function specifies that the input bit-width should be between 32 and 52 on Line 2 (i.e., \emph{pre-conditions}) and that the return value of this function must be less than $2^{52}$ or $0x10\_0000\_0000\_0000$ on Line 3 (\emph{post-condition}). All the callers of \texttt{MAX\_PHYADDR} function need to verify that the arguments they pass to \texttt{MAX\_PHYADDR} always satisfy the pre-condition. 
Inside \texttt{MAX\_PHYADDR}, we need to verify that the post-condition is guaranteed to hold given the pre-condition. To achieve this goal,
developers provided \emph{proof annotations} on Line 5--8 to aid the Verus verifier in proving those obligations.
In this case, the proof annotations tell Verus to use bit-vector reasoning to prove that line 9 will not underflow
and that it will satisfy the post-condition.
To verify the function, Verus turns the executable code, specification, and proof annotations into a query for Z3 \cite{z3}, Verus's underlying SMT solver.

\begin{figure}
\footnotesize{
    \centering
    \begin{minipage}{0.62\linewidth}
    \begin{lstlisting}[language=Verus,style=paperlisting,escapechar=\%]
pub fn MAX_PHYADDR(max_width:u64) -> ( ret : u64)
  requires 32 <= max_width <= 52,
  ensures ret < 0x10_0000_0000_0000u64,
{
  assert(1u64 << max_width > 1) by(bit_vector)
    requires 32 <= max_width <= 52;
  assert(1u64 << max_width <= 0x10_0000_0000_0000u64) by(bit_vector)
    requires 32 <= max_width <= 52;
  (1u64 << max_width) - 1u64
}
\end{lstlisting}
    \end{minipage}
\caption{A Verus-verified function simplified from NRKernel}
\label{fig:example}
}
\end{figure}
\section{Full VeruSAGE Agent Design \label{app:agent}\label{sec:design}}

In building/using agentic systems to support LLMs on proof tasks, we take two different approaches, which we call \emph{hands-off} (\S\ref{sec:hands-off}) and \emph{hands-on} (\S\ref{sec:hands-on}).

\subsection{Hands-Off Approach}
\label{sec:hands-off}

The hands-off approach uses a generic coding agent and a simple prompt listed below, which focuses on warning the LLM not to cheat, with little information about verification in general or Verus in particular. The LLM is merely given the option of (1) running Verus, (2) running a Verus cheat checker that reports whether a proof features cheating, and (3) inspecting the contents of a folder containing the Verus standard library (vstd).

\begin{figure}[h]
\vspace{-0.2in}
\begin{lstlisting}[breaklines=true, breakindent=0pt, language=, numbers = none, escapechar=\%,]

The file X.rs cannot be verified by Verus, a veri-fication tool for Rust programs, yet. Please add proof annotations to X.rs so that it can be successfully verified by Verus, and write the resulting code with proof into a new file, X_verified.rs. Please invoke Verus to check the proof annotation you added. The vstd folder in the current directory is a copy of Verus' vstd definitions and helper lemmas; please feel free to check it when needed. You should KEEP editing your proof annotations until Verus shows there is no error. You should NOT change existing functions' pre-conditions or post-conditions; you should NOT change any executable Rust code; and you should NEVER use admit(...) or assume(...) in your code. You are also NOT allowed to create unimplemented, external-body lemma functions --- for any new lemma functions you add, you should provide complete proof. You are NOT allowed to create new axiom functions or change the pre/post conditions of existing axiom functions, and you should NEVER add external_body tag to any existing non-external-body functions. I have installed Verus locally; you can just run Verus. Before you are done, MAKE SURE to run verus-checker X_verified.rs to double check whether you have made any illegal changes to X.rs (fix those if you did).
\end{lstlisting}
\vspace{-0.1in}
\end{figure}

\MyPara{Setup} To run this hands-off approach, we use GitHub Copilot Command-Line Interface (CLI)~\cite{github_copilot_cli} for all the models that it supports (\gfive, \sfour, \sfourfive, \opusfourfive), and OpenAI Codex CLI~\cite{openai_codex} for \gmini. We use Copilot's \codeIn{allow-all-tools} and Codex's \codeIn{all-auto} option, so that the agent can run Verus or any command-line tools without prompting human users for confirmation. In the remainder of the paper, we refer to all these agent systems as \textit{hands-off} and/or as \textit{CLI agents}.

To make sure that the LLM does not find ``answers'' somewhere in the file system, we run all experiments in containers. When we evaluate tasks from a project $P$, we ensure that no proof bodies from $P$ exist in that container.
Each run of a CLI agent produces two files: its standard output log and the verified file. As we will see later in Figure \ref{fig:mono}, the CLI log is very informative, indicating which files the agent reads, what commands, including Verus, it executes, etc. 

\subsection{Hands-On Approach}
\label{sec:hands-on}

A hands-on approach, in contrast, prompts the LLM with detailed and comprehensive domain knowledge about Verus syntax, verification-error debugging strategies, and various other support. It also guides and enforces a proof development methodology, rather than relying on the LLM to drive the development process through self-driven tool invocations.

We do not use the state-of-the-art hands-on approach \autoverus~\cite{autoverus} directly, as it performs poorly on system tasks. As shown in Table~\ref{tab:autoverus-verusysbench}, it has only 20\% success rate on \bench, matching the findings in RagVerus~\cite{rvbench}.

To improve AutoVerus for system tasks, we did a deep dive on two projects, Memory Allocator (MA) and Storage (ST). We interviewed their authors and examined their human-written proof to learn how human experts developed the proofs therein and what parts of this methodology are missing from the \autoverus agent system.
This understanding then led to the following design of \tool. (Note that the design of \tool is \textbf{not} influenced by the benchmark study presented in \S\ref{sec:bench-analysis}. MA and ST were the first two system projects that we looked into and extracted proof tasks from; most of the study and benchmark extraction presented in \S\ref{sec:bench-analysis} was conducted after the design of \tool.)

\begin{figure}
    \centering
    \includegraphics[width=0.9\linewidth]{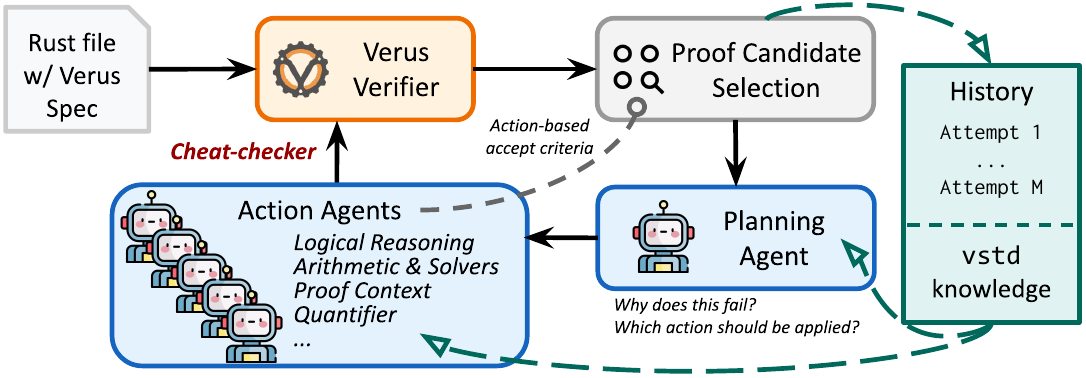}
    \caption{Architecture of \tool}
    \label{fig:handson}
\end{figure}

As illustrated in Figure~\ref{fig:handson}, for every task, the proof-development of \tool goes through \textit{steps}. Each step starts with Verus reporting verification errors, if any, of the Rust program with the most up-to-date proof candidate; the candidate selector then decides whether to take this proof candidate or revert to an earlier one; then, the planning agent examines the errors and the context history to select an action agent, which then proposes an updated proof candidate. \tool 
differs from \autoverus with (1) many more action agents; (2) a two-phase plan-then-act LLM procedure; (3) more sophisticated candidate selector; and (4) more context management.

\MyPara{What action agents are missing in AutoVerus?}
AutoVerus contains a network of action agents, each designed to refine loop invariants in a particular way or to fix one type of verification errors, such as pre-/post-condition errors and loop-invariant errors. However, our analysis reveals that AutoVerus' error-driven agents lack knowledge of high-level proof strategies (e.g., divide-and-conquer, induction), specialized solvers (e.g., bit-vector and nonlinear solvers), and other verification techniques beyond those for loops. 
To support this, \verusys is equipped with many more action agents, including: \textit{Logical Reasoning} agents (e.g., \codeIn{case-analysis}, \codeIn{induction}), \textit{Arithmetic \& Solvers} agents (e.g., \codeIn{nonlinear-arithmetic}, \codeIn{bit-vector}, \codeIn{integer-ring}), \textit{Proof Context} agents (e.g., \codeIn{reveal-opaque}, \codeIn{use-lemma}), \textit{Quantifier} agents (e.g., \codeIn{instantiate-forall}, \codeIn{instantiate-exists}), etc.

\MyPara{Why do LLMs need to plan before act?}
\autoverus does not have a planning phase, as it simply dispatches an action agent based on the verification error type.
In more complicated proof tasks, one type of error might be fixed by different strategies/techniques. For example, \tool has 16 specialized action agents to help fix assertion errors. One needs to analyze the code context and failure history to select the most promising strategy or technique, rather than blindly trying one. 
To support this, \tool has a dedicated planning agent that is equipped with the high-level description of every action agent, as well as a when-to-use-what tutorial. 

Furthermore, to help the planning agent, \tool performs a static analysis of the codebase before invoking the planning agent to identify available resources such as lemmas, recursive functions, and opaque functions. Based on this analysis, \tool filters the available actions to exclude those that are not applicable (e.g., disabling \codeIn{use-lemma} if no lemmas are present in the task file) and prioritize relevant ones (e.g., boosting agents that are good at handling recursion if recursive functions are detected). This reduces the search space and guides the model towards more promising actions.

\MyPara{When do we accept a proof candidate?}
Before the proof is finalized, many versions of intermediate proof, referred to as \textit{proof candidates}, are usually proposed and judged by the verifier as incorrect. It is important to identify which proof candidates have made useful progress and accept them for further exploration. 
In simple tasks, this can be easily decided by identifying the proof candidate with the fewest verification errors --- this is the strategy of \autoverus. Unfortunately, this strategy does not work for complicated proof tasks. For example, converting one difficult-to-prove property into multiple easier properties would temporarily increase the number of errors but is often the right way to move forward.

\verusys proposes more types of acceptance criteria and uses different criteria for different actions. For example, its \codeIn{case-analysis} action implements the divide-and-conquer strategy by splitting a proof into multiple branches. \tool applies an acceptance criterion that allows the number of verification errors to increase as long as the original assertion error targeted by this action is resolved and no errors are introduced outside the split block. In contrast, for actions like \codeIn{nonlinear-arithmetic}, which invokes a specialized solver to prove a specific arithmetic property, we enforce a stricter criterion: a candidate is accepted only if it strictly reduces the total number of verification errors.

\MyPara{Context management.} \tool aims to provide sufficient context information to the LLM planning agent in several ways.
(1) \textit{Comprehensive History Tracking}: \tool maintains a log of past proof attempts, recording what action agent was used, what is the synthesized code diff, and what is the outcome --- whether the proof candidate was accepted, what are the verification errors, etc. This log is included in the prompt for subsequent invocations of the planning agent, allowing the agent to learn from past mistakes and refine its overall strategy.
(2) \textit{Concise Code Context}: Not to overwhelm the model with the entire codebase, we provide a focused view containing only the relevant code surrounding the failure and the specific error message. Furthermore, to minimize token usage and focus the model on the specific changes, we instruct action agents to output proof candidates as concise code diffs (specifically, search-and-replace blocks) rather than rewriting entire files.
Crucially, this context is \textit{dynamic}: the history evolves, allowing the planning to adapt its strategy in real-time and avoid repeating past mistakes.

\MyPara{Setup.} We implemented \tool by adding about 15,600 lines of Python code into \autoverus for the design changes mentioned above. \tool uses Azure OpenAI APIs to access OpenAI and Claude models directly. 

To ensure a controlled evaluation, we enforce a stopping criterion: \tool terminates only when (1) the code is successfully verified by Verus, (2) the total execution time exceeds 20 minutes, or (3) 20 proof steps have been conducted. This policy ensures the agent has sufficient opportunity to explore complex strategies but does not run indefinitely. 
\section{Extended Experimental Results}
\label{app:extended-eval}

This appendix contains material deferred from \S\ref{sec:eval} due to the page limit: full per-project per-configuration tables, case study on how/whether LLMs can help finish proofs that human experts have not yet finished, ablations on alternative agent settings, the Verus/cheat-checker ablation, additional case studies (multi-file projects, no-lemma variant, model collaboration, multi-run reliability), results of the open-weight Qwen3.5, and statistical analyses.

\subsection{Full per-project results (Hands-Off and Hands-On)}

\begin{table}[h]
    \caption{Detailed Hands-\textbf{Off} results per project (time, cost, and Verus runs are averages per task; S4/S4.5: Sonnet 4/4.5; \colorbox{rankone}{best} and \colorbox{rankfour}{worst} highlighted; \colorbox{rankone}{\bf bold}: better than the best in Table~\ref{tab:details-verusys}).}
    \label{tab:details-cli}
    \tablestyle
    \scriptsize
    \setlength{\tabcolsep}{3pt}
    \begin{tabular}{lrrrrrrrrrrrrrrrr}
    \toprule
            & \multicolumn{4}{c}{Success rate}
            & \multicolumn{4}{c}{Time (min)}
            & \multicolumn{4}{c}{Cost (\$)}
            & \multicolumn{4}{c}{Verus runs} \\
    \cmidrule(lr){2-5} \cmidrule(lr){6-9} \cmidrule(lr){10-13} \cmidrule(lr){14-17}
    Project & \gmini & \gfive & S4 & S4.5 & \gmini & \gfive & S4 & S4.5 & \gmini & \gfive & S4 & S4.5 & \gmini & \gfive & S4 & S4.5 \\
    \midrule
    AL     &\cellcolor{rankfour}25\%&\cellcolor{rankthree}77\%&\cellcolor{ranktwo}86\%&\cellcolor{rankone}\textbf{100\%}&\cellcolor{rankthree}3.8&\cellcolor{ranktwo}3.3&\cellcolor{rankfour}3.9&\cellcolor{rankone}\textbf{3.1}&\cellcolor{ranktwo}1.00&\cellcolor{rankone}0.81&\cellcolor{rankfour}2.57&\cellcolor{rankthree}1.86&10.4&6.6&9.0&7.0\\
    AC     &\cellcolor{rankfour}11\%&\cellcolor{rankthree}16\%&\cellcolor{ranktwo}24\%&\cellcolor{rankone}\textbf{37\%}&\cellcolor{rankone}\textbf{2.1}&\cellcolor{ranktwo}2.7&\cellcolor{rankfour}14.5&\cellcolor{rankthree}12.5&\cellcolor{rankone}\textbf{0.57}&\cellcolor{ranktwo}1.96&\cellcolor{rankfour}15.80&\cellcolor{rankthree}12.31&2.6&4.8&17.0&13.6\\
    IR     &\cellcolor{rankfour}25\%&\cellcolor{rankthree}62\%&\cellcolor{ranktwo}69\%&\cellcolor{rankone}\textbf{84\%}&\cellcolor{ranktwo}4.3&\cellcolor{rankone}\textbf{3.9}&\cellcolor{rankfour}7.4&\cellcolor{rankthree}6.7&\cellcolor{rankone}0.91&\cellcolor{ranktwo}1.09&\cellcolor{rankfour}5.10&\cellcolor{rankthree}4.56&10.7&6.3&10.3&9.5\\
    MA     &\cellcolor{rankfour}15\%&\cellcolor{rankthree}60\%&\cellcolor{ranktwo}75\%&\cellcolor{rankone}\textbf{90\%}&\cellcolor{ranktwo}4.9&\cellcolor{rankone}\textbf{3.2}&\cellcolor{rankfour}8.3&\cellcolor{rankthree}5.2&\cellcolor{rankone}0.85&\cellcolor{ranktwo}1.14&\cellcolor{rankfour}6.00&\cellcolor{rankthree}3.81&9.8&8.2&11.6&9.6\\
    NO     &\cellcolor{rankfour}48\%&\cellcolor{ranktwo}93\%&\cellcolor{rankthree}86\%&\cellcolor{rankone}\textbf{100\%}&\cellcolor{rankthree}3.9&\cellcolor{ranktwo}2.6&\cellcolor{rankfour}4.5&\cellcolor{rankone}2.0&\cellcolor{rankthree}1.17&\cellcolor{rankone}0.46&\cellcolor{rankfour}2.18&\cellcolor{ranktwo}0.85&11.3&4.5&7.6&4.9\\
    NR     &\cellcolor{rankfour}14\%&\cellcolor{rankthree}39\%&\cellcolor{ranktwo}55\%&\cellcolor{rankone}\textbf{74\%}&\cellcolor{rankone}\textbf{2.9}&\cellcolor{ranktwo}3.7&\cellcolor{rankfour}12.3&\cellcolor{rankthree}8.9&\cellcolor{rankone}0.78&\cellcolor{ranktwo}1.48&\cellcolor{rankthree}4.50&\cellcolor{rankfour}6.35&6.2&5.9&11.2&11.0\\
    OS     &\cellcolor{rankfour}9\%&\cellcolor{rankthree}43\%&\cellcolor{ranktwo}62\%&\cellcolor{rankone}\textbf{83\%}&\cellcolor{rankone}\textbf{2.4}&\cellcolor{ranktwo}3.8&\cellcolor{rankfour}9.9&\cellcolor{rankthree}8.3&\cellcolor{rankone}0.66&\cellcolor{ranktwo}1.33&\cellcolor{rankthree}6.79&\cellcolor{rankfour}7.15&6.0&5.6&9.6&12.8\\
    ST     &\cellcolor{rankfour}9.5\%&\cellcolor{rankthree}33\%&\cellcolor{ranktwo}70\%&\cellcolor{rankone}\textbf{78\%}&\cellcolor{rankone}\textbf{2.0}&\cellcolor{ranktwo}4.9&\cellcolor{rankfour}9.1&\cellcolor{rankthree}7.4&\cellcolor{rankone}\textbf{0.42}&\cellcolor{ranktwo}1.57&\cellcolor{rankfour}5.90&\cellcolor{rankthree}5.84&2.1&8.0&10.5&9.9\\
    VE     &\cellcolor{rankfour}27\%&\cellcolor{rankthree}64\%&\cellcolor{ranktwo}82\%&\cellcolor{rankone}\textbf{100\%}&\cellcolor{ranktwo}4.3&\cellcolor{rankfour}5.7&\cellcolor{rankthree}5.2&\cellcolor{rankone}\textbf{2.4}&\cellcolor{rankone}1.32&\cellcolor{rankthree}1.51&\cellcolor{rankfour}4.00&\cellcolor{ranktwo}1.50&13.3&10.0&10.9&7.1\\
    \midrule
    All    &\cellcolor{rankfour}17\%&\cellcolor{rankthree}50\%&\cellcolor{ranktwo}64\%&\cellcolor{rankone}\textbf{81\%}&\cellcolor{rankone}\textbf{3.3}&\cellcolor{ranktwo}3.7&\cellcolor{rankfour}9.2&\cellcolor{rankthree}7.2&\cellcolor{rankone}0.81&\cellcolor{ranktwo}1.30&\cellcolor{rankfour}6.65&\cellcolor{rankthree}5.61&7.4&6.4&10.9&10.4\\
    \bottomrule
    \end{tabular}
\end{table}

\begin{table}[h]
    \caption{Detailed Hands-\textbf{On} results per project (time, cost, and Verus runs are averages per task; S4/S4.5: Sonnet 4/4.5; \colorbox{rankone}{best} and \colorbox{rankfour}{worst} highlighted; \colorbox{rankone}{\bf bold}: better than the best in Table~\ref{tab:details-cli}).}
    \label{tab:details-verusys}
    \tablestyle
    \scriptsize
    \setlength{\tabcolsep}{3pt}
    \begin{tabular}{lrrrrrrrrrrrrrrrr}
    \toprule
            & \multicolumn{4}{c}{Success rate}
            & \multicolumn{4}{c}{Time (min)}
            & \multicolumn{4}{c}{Cost (\$)}
            & \multicolumn{4}{c}{Verus runs} \\
    \cmidrule(lr){2-5} \cmidrule(lr){6-9} \cmidrule(lr){10-13} \cmidrule(lr){14-17}
    Project & \gmini & \gfive & S4 & S4.5 & \gmini & \gfive & S4 & S4.5 & \gmini & \gfive & S4 & S4.5 & \gmini & \gfive & S4 & S4.5 \\
    \midrule
    AL     &\cellcolor{rankfour}48\%&\cellcolor{ranktwo}79\%&\cellcolor{rankthree}69\%&\cellcolor{rankone}83\%&\cellcolor{rankfour}12.4&\cellcolor{rankthree}8.7&\cellcolor{ranktwo}6.7&\cellcolor{rankone}4.6&\cellcolor{ranktwo}0.45&\cellcolor{rankone}\textbf{0.31}&\cellcolor{rankfour}0.60&\cellcolor{rankthree}0.51&32.9&10.0&25.4&20.3\\
    AC     &\cellcolor{rankfour}19\%&\cellcolor{ranktwo}32\%&\cellcolor{rankthree}21\%&\cellcolor{rankone}\textbf{37\%}&\cellcolor{rankfour}16.9&\cellcolor{rankthree}16.5&\cellcolor{ranktwo}14.6&\cellcolor{rankone}13.1&\cellcolor{ranktwo}1.35&\cellcolor{rankone}0.86&\cellcolor{rankfour}6.09&\cellcolor{rankthree}5.20&36.8&13.3&31.6&25.5\\
    IR     &\cellcolor{rankfour}35\%&\cellcolor{rankthree}44\%&\cellcolor{ranktwo}53\%&\cellcolor{rankone}67\%&\cellcolor{rankthree}13.8&\cellcolor{rankfour}14.1&\cellcolor{ranktwo}7.5&\cellcolor{rankone}7.1&\cellcolor{ranktwo}0.54&\cellcolor{rankone}\textbf{0.48}&\cellcolor{rankfour}1.03&\cellcolor{rankthree}0.92&45.7&17.0&35.3&28.5\\
    MA     &\cellcolor{rankfour}62\%&\cellcolor{rankthree}72\%&\cellcolor{ranktwo}75\%&\cellcolor{rankone}84\%&\cellcolor{rankthree}9.9&\cellcolor{rankfour}10.2&\cellcolor{ranktwo}4.4&\cellcolor{rankone}3.9&\cellcolor{ranktwo}0.37&\cellcolor{rankone}\textbf{0.35}&\cellcolor{rankfour}0.53&\cellcolor{rankthree}0.40&30.9&13.2&26.5&20.3\\
    NO     &\cellcolor{rankfour}72\%&\cellcolor{rankthree}83\%&\cellcolor{rankone}\textbf{100\%}&\cellcolor{ranktwo}97\%&\cellcolor{rankfour}9.2&\cellcolor{rankthree}7.6&\cellcolor{rankone}\textbf{1.2}&\cellcolor{ranktwo}1.6&\cellcolor{rankfour}0.29&\cellcolor{rankthree}0.25&\cellcolor{ranktwo}0.19&\cellcolor{rankone}\textbf{0.18}&31.0&9.4&25.1&19.4\\
    NR     &\cellcolor{rankfour}30\%&\cellcolor{rankthree}48\%&\cellcolor{ranktwo}56\%&\cellcolor{rankone}60\%&\cellcolor{rankfour}14.0&\cellcolor{rankthree}13.2&\cellcolor{rankone}7.8&\cellcolor{ranktwo}8.2&\cellcolor{ranktwo}0.79&\cellcolor{rankone}\textbf{0.60}&\cellcolor{rankfour}2.01&\cellcolor{rankthree}1.84&41.0&13.4&32.3&26.3\\
    OS     &\cellcolor{rankfour}37\%&\cellcolor{rankthree}45\%&\cellcolor{ranktwo}54\%&\cellcolor{rankone}62\%&\cellcolor{rankthree}12.7&\cellcolor{rankfour}14.6&\cellcolor{rankone}7.0&\cellcolor{ranktwo}7.7&\cellcolor{ranktwo}0.81&\cellcolor{rankone}\textbf{0.62}&\cellcolor{rankfour}2.10&\cellcolor{rankthree}1.69&36.1&14.4&30.6&24.3\\
    ST     &\cellcolor{rankfour}49\%&\cellcolor{ranktwo}62\%&\cellcolor{rankthree}54\%&\cellcolor{rankone}71\%&\cellcolor{rankthree}11.2&\cellcolor{rankfour}12.7&\cellcolor{rankone}5.4&\cellcolor{ranktwo}5.9&\cellcolor{ranktwo}0.53&\cellcolor{rankone}0.51&\cellcolor{rankfour}1.31&\cellcolor{rankthree}1.03&31.6&13.0&29.6&23.4\\
    VE     &\cellcolor{rankthree}68\%&\cellcolor{ranktwo}73\%&\cellcolor{rankfour}64\%&\cellcolor{rankone}77\%&\cellcolor{rankthree}9.0&\cellcolor{rankfour}11.7&\cellcolor{rankone}3.8&\cellcolor{ranktwo}4.5&\cellcolor{rankone}\textbf{0.38}&\cellcolor{ranktwo}0.48&\cellcolor{rankfour}0.65&\cellcolor{rankthree}0.58&46.1&20.5&34.2&25.8\\
    \midrule
    All    &\cellcolor{rankfour}41\%&\cellcolor{rankthree}56\%&\cellcolor{ranktwo}58\%&\cellcolor{rankone}67\%&\cellcolor{rankfour}12.8&\cellcolor{rankthree}12.7&\cellcolor{ranktwo}7.1&\cellcolor{rankone}6.9&\cellcolor{ranktwo}0.67&\cellcolor{rankone}\textbf{0.52}&\cellcolor{rankfour}1.72&\cellcolor{rankthree}1.47&37.7&13.7&30.3&24.1\\
    \bottomrule
    \end{tabular}
\end{table}

\begin{table}[h]
    \caption{\opusfourfive Hands-Off results per project (time and cost are averages per task).}
    \label{tab:details-opus}
    \tablestyle
    \begin{tabular}{lrrrr}
    \toprule
    Project & \#Tasks & Success rate & Time (min) & Cost (\$) \\
    \midrule
    AL  & 104 & 100\% & 1.3  & 1.86  \\
    AC  & 63  & 83\%  & 22.8 & 32.87 \\
    IR  & 118 & 96\%  & 7.6  & 9.78  \\
    MA  & 89  & 94\%  & 3.4  & 5.53  \\
    NO  & 29  & 100\% & 1.3  & 2.21  \\
    NR  & 204 & 91\%  & 32.0 & 22.30 \\
    OS  & 157 & 93\%  & 18.4 & 15.60 \\
    ST  & 63  & 97\%  & 5.3  & 9.12  \\
    VE  & 22  & 100\% & 1.4  & 1.88  \\
    \midrule
    All & \tasks & 94\% & 15.3 & 13.65 \\
    \bottomrule
    \end{tabular}
\end{table}

\subsection{Alternative Verus agent system and frontier models on other benchmarks}

Table~\ref{tab:autoverus-verusysbench} reports AutoVerus's success rate per project; the 20\% overall rate matches the findings reported by RagVerus~\cite{rvbench}.

\begin{table}[h]
    \caption{Success rate of AutoVerus on each project.}
    \label{tab:autoverus-verusysbench}
    \tablestyle
    \begin{tabular}{lrrrrrrrrrr}
    \toprule
     & AL & AC & IR & MA & NO & NR & OS & ST & VE & All \\
    \midrule
    Success rate & 30\% & 6\% & 24\% & 32\% & 34\% & 15\% & 13\% & 19\% & 23\% & 20\% \\
    \bottomrule
    \end{tabular}
\end{table}

Several previous projects \cite{autoverus, exverus, alphaverus, safeiclr25} have reported high success rates of models generating proofs for VerusBench tasks. However, there has been no report on how successful models are in generating proofs for HumanEval-Verus tasks. We extracted 74 proof tasks from HumanEval-Verus, together with 27 no-lemma tasks. We ran \sfourfive using the Hands-Off setting. We see that \sfourfive
achieves 92\% success rate on the 74 normal tasks, and 74\% success rate on the no-lemma tasks. Compared with \sfourfive's performance on \bench (81\% on normal tasks and 57\% on no-lemma tasks), this confirms our intuition that system proof is more difficult than small-scale algorithmic task proof.

\subsection{Can LLMs help with tasks not yet finished by humans?}
\label{sec:eval-help-experts}

We evaluate whether the capable combination (Hands-Off \sfourfive) can help with proof tasks human experts have \textbf{not} yet finished. Atmosphere (OS) offers this opportunity: being very recent, some of it is still under proof development.

We first consider Atmosphere's \verb|lemma_u.rs| file. Among its 13 lemmas, 10 are unproved. After 12 minutes at a cost of \$11 (USD), \sfourfive produces a fully verified file. Notably, \sfourfive points out five lemmas whose specification requires adjustment, all five confirmed by human experts. Figure~\ref{fig:seqskip} shows one such adjustment. The original specification of \verb|seq_skip_lemma| states that for any non-empty sequence \verb|s|, once its first element \verb|s[0]| is skipped, the remaining sequence contains no element whose value equals that of the original \verb|s[0]|. As the log indicates (Figure~\ref{fig:mono}), after multiple proof attempts rejected by Verus, \sfourfive starts to doubt the correctness of this specification and proposes the correct adjustment. OS is not the only project where LLMs propose useful specification adjustments: we have seen another example in ST accepted upstream~\cite{storagepull}.

\UseRawInputEncoding
\begin{figure}
\centering
\begin{minipage}{0.62\linewidth}
\begin{lstlisting}[language=Verus,style=paperlisting,escapechar=\%,numbers=none,xleftmargin=0pt]
pub proof fn seq_skip_lemma<A>()
  ensures
    forall|s: Seq<A>| s.len() > 0
%\Added%+     && s.no_duplicates() ==> !s.skip(1).contains(s[0]),
\end{lstlisting}
\end{minipage}
\caption{A specification adjustment suggested by an LLM (the added conjunct is shaded).}
\label{fig:seqskip}
\end{figure}
\UseRawInputEncoding

\begin{figure}
\centering
\begin{minipage}{0.8\linewidth}
\begin{lstlisting}[language=,style=paperlisting,breaklines=true,breakindent=0pt,numbers=none,xleftmargin=0pt,escapechar=\%]

%$\bullet$% Wait, let me reconsider. Looking at the post-condition again - actually it might be wrong. Let me check the original file to see if this postcondition is needed elsewhere. ...
%$\bullet$% I see - the postcondition says `s.len() > 0 ==> !s.skip(1).contains(s[0])`. This can`t be true in general! If s = [1, 1], then s.skip(1) = [1] which contains s[0] = 1. Let me think... Maybe this postcondition needs no_duplicates too? Or maybe I`m misunderstanding skip(1). Let me check the implementation of skip more carefully and test:
%\checkmark% Check skip definition
   $ cd /xxx/verified && grep -A3 "fn skip" vstd/seq.rs | head -10
%$\bullet$% So skip(1) = subrange(1, len) which is the tail of the sequence. Now I see-this DOES need no_duplicates!
\end{lstlisting}
\end{minipage}
    \caption{The monologue of \sfourfive for \codeIn{seq\_skip\_lemma}.}
\label{fig:mono}
\end{figure}

In other parts of Atmosphere, we find 25 functions with specifications but incomplete proofs: 8 have empty bodies intended to be completed later; 17 contain substantial proof material but use one or two \verb|assume|s. We extract these as tasks. \sfourfive generates complete proofs for 23 of them in 13.5 minutes and \$8.45 on average.

Importantly, humans and \sfourfive are effective \textit{collaborators}. For the 17 incomplete human proofs, \sfourfive proves 16 when given the partial human proof but only 6 starting from scratch; even on those 6, starting from the partial proof cuts average development time from 7.3 to 4.7 minutes.

\subsection{What about alternative settings?}
\label{sec:eval-alternate}

\MyPara{The Hands-On approach does not suit every model.} \tool helped \gmini significantly and \gfive somewhat, but hurt Sonnet models relative to Hands-Off. One indicator: in Hands-Off mode the final proof contains syntax errors in 38\% of \gmini tasks and 16\% of \gfive tasks, but only 1.3\% of \sfour and 0.9\% of \sfourfive tasks. \gmini and \gfive benefit from \tool's tutorial and syntax-repair agents; Sonnet does not need them.

A likely reason \tool hurts Sonnet is that it forces small, per-error steps. In \codeIn{segment\_start\_mult\_commit\_size} (MA), when \sfourfive finally found the correct proof after 10 \tool steps, it was already past the 20-minute cut-off and counted as a failure. Hands-Off \sfourfive proposed a 53-line change to the 52-line input file in 3 tries over 2.5 minutes and finished.

Occasionally, \tool's hard-coded policies override a better choice the model would have made: the candidate selector cannot always measure an incomplete proof's potential, and the static-analysis cheat checker can produce false positives that reject correct proofs.

\MyPara{Even Hands-Off LLMs benefit from tool support.}
First, LLMs need Verus feedback. Even in Hands-Off, \gfive averages 6.4 Verus runs per task. \sfourfive starts every task by running Verus on the input file; \gfive often writes a proof before seeing any Verus output. \sfourfive used only one Verus run in just 21 tasks (3\% of its successes); once, it ran Verus 50 times before succeeding.

Second, the cheat checker reduces cheating. Without it, \sfour cheated on 14\% of tasks; \sfourfive and \gfive cheated on 7\% and 2\% respectively. With the cheat checker in the prompt, all models' cheat rates fall below 1.5\%. The residual cheats tend to happen when the model cannot find a proof.

Finally, a full ablation of Hands-Off without \codeIn{vstd} and without the cheat checker shows that together they enable \gfive, \sfour, and \sfourfive to prove 26\%, 24\%, and 9\% more \bench tasks respectively. Models use \codeIn{vstd} to find relevant helper lemmas and also to learn Verus syntax. The support increases time per task by 53\% for \gfive, 21\% for \sfour, and 0.4\% for \sfourfive.

\subsection{More detailed investigations}
\label{sec:eval-detailed-investigations}

\MyPara{No-lemma variant.}
\label{sec:nolemma}
The harder \textit{no-lemma} variant of \bench (see \S\ref{sec:bench-setup}) removes the helper-lemma stubs from each task file; this applies to the \noltasks of \tasks tasks whose human-written proofs use at least one helper lemma. Table~\ref{tab:nolemma} reports Hands-Off success, time, and cost per project on this variant for the two strongest models, \sfourfive and \opusfourfive.

\begin{table}[h]
    \caption{Hands-Off results on the \textit{no-lemma} variant of \bench{} (time and cost are averages per task; S4.5/O4.5: Sonnet 4.5/Opus 4.5; the better of the two models in bold).}
    \label{tab:nolemma}
    \tablestyle
    \begin{tabular}{lrrrrrrr}
    \toprule
            & & \multicolumn{2}{c}{Success rate}
            & \multicolumn{2}{c}{Time (min)}
            & \multicolumn{2}{c}{Cost (\$)} \\
    \cmidrule(lr){3-4} \cmidrule(lr){5-6} \cmidrule(lr){7-8}
    Project & \#Tasks & S4.5 & O4.5 & S4.5 & O4.5 & S4.5 & O4.5 \\
    \midrule
    AL & 83 & 95\% & \textbf{100\%} & 2.6 & \textbf{1.8} & 2.7 & \textbf{1.5} \\
    AC & 58 & 19\% & \textbf{41\%} & \textbf{10.7} & 45.9 & \textbf{11.5} & 96.3 \\
    IR & 30 & 50\% & \textbf{97\%} & 11.7 & \textbf{10.1} & \textbf{13.2} & 21.5 \\
    MA & 47 & 72\% & \textbf{100\%} & \textbf{9.9} & 25.2 & \textbf{11.8} & 24.4 \\
    NO & 5 & \textbf{100\%} & \textbf{100\%} & 5.7 & \textbf{3.2} & \textbf{6.3} & 7.3 \\
    NR & 133 & 44\% & \textbf{84\%} & \textbf{14.2} & 46.4 & \textbf{14.1} & 85.2 \\
    OS & 71 & 58\% & \textbf{96\%} & \textbf{15.5} & 45.6 & \textbf{12.2} & 59.2 \\
    ST & 29 & 55\% & \textbf{93\%} & \textbf{10.5} & 14.9 & \textbf{10.9} & 28.0 \\
    VE & 4 & \textbf{100\%} & \textbf{100\%} & 6.0 & \textbf{2.6} & \textbf{3.6} & \textbf{3.6} \\
    \midrule
    All & \noltasks & 57\% & \textbf{87\%} & \textbf{10.9} & 30.8 & \textbf{10.8} & 52.0 \\
    \bottomrule
    \end{tabular}
\end{table}

\opusfourfive remains highly capable without lemma scaffolding (\textbf{87\%} overall, vs.~92\% with lemmas on the same \noltasks tasks: a 5-point drop), demonstrating that the strongest model can re-derive most intermediate facts on its own. \sfourfive degrades much more sharply (57\% vs.~71\% on the same tasks: a 14-point drop), which suggests it relies more heavily on the human-provided lemma structure. The biggest absolute drops for both models concentrate in AC, NR, OS, ST, and IR --- projects whose human proofs depend on long chains of helper lemmas that become expensive to re-derive.

The no-lemma variant is also markedly more expensive in time and money. \opusfourfive averages \textbf{31~minutes and \$52} per task (vs.\ \opustimeoverall and \opuscostoverall with lemmas: $\sim$2$\times$ slower and $\sim$4$\times$ more expensive), while \sfourfive averages \textbf{11~minutes and \$11} per task (vs.\ \sftimeoverall and \sfcostoverall: $1.5\times$ slower, $\sim$2$\times$ more expensive). The cost blow-up concentrates in the projects with the largest success drops (AC, NR, OS), where the model spends many turns on long inductive chains before either succeeding or giving up; on AC alone, \opusfourfive averages \$96 per task.

\MyPara{Multi-file project setting.}
Throughout the main study, we extract each proof task into a stand-alone file. We also tested whether LLMs can work directly in the multi-file project. On three IronKV tasks (with all local proofs stripped), \sfourfive (Hands-Off) proved all three in the whole-project setting, taking 1.6$\times$ the time and 2.2$\times$ the cost of the single-file setting. On Atmosphere (10 folders, 150+ files), we ran two tasks that single-file \sfourfive solved: one verified (1.4$\times$ time, 1.8$\times$ cost); the other, \codeIn{add\_mapping\_4k}, exceeded Copilot's token quota after 1 hour and over \$40 (13.8M tokens) without finishing. For now, extracting dependencies into one file is the preferred route. Note that the whole-project setting is not always harder: models need to find the code they depend on, but they may also see and imitate other proofs in the project. For example, \sfourfive completes 16 of 17 incomplete Atmosphere proofs when given the partial human proofs, but only 6 of 17 from scratch (Appendix~\ref{sec:eval-help-experts}). Our repo-level track (\S\ref{sec:bench-setup}) puts every task back into its original repository, so that such effects can be studied on all tasks.

\MyPara{Model collaboration.}
``Union'' of all four models in Hands-Off reaches 82.3\%, only 1.4 points above Sonnet 4.5 alone. A handoff strategy -- cheap \gmini does initial steps, then \sfourfive refines -- did not lower cost on two MA tasks: \sfourfive starting from \gmini's partial proof took 20 min (\$13.70) and 39 min (\$18.20), whereas from scratch it took 2 min (\$0.81) and 16 min (\$14.70).

\MyPara{Multi-run reliability.}
A second run of Hands-Off \sfourfive succeeds on $\sim$10\% of initial failures in AC, NR, and ST. Running \tool + \gmini three times on 100 random tasks: accumulated success rate grows from 41\% (run 1) to 56\% (run 3), but each individual run's success rate stays near 40\% (41\%, 39\%, 39\%) -- the gains come from different tasks, not reliability.

\subsection{Open-weight model: Qwen3.5}
\label{app:qwen}

To see how open models perform on \bench, we evaluated Qwen3.5 (397B-A17B, FP8, with Apache-2.0 weights)~\cite{qwen35} on all \tasks normal tasks under both agent settings. We served the model ourselves with vLLM. For Hands-Off, we used OpenAI Codex CLI connected to the vLLM server (the same agent as for \gmini) with the same prompt as the other models. To avoid runaway sessions, we stopped each session after 45 minutes, and when the model ended a session without writing an output file, we resumed the same session up to three times. For Hands-On, we used \tool with its usual limit of 20 steps or 20 minutes. Everything else (Verus version, cheat checker, container setup) is the same as in the main study, and any output flagged by the cheat checker is counted as a failure.

\begin{table}[h]
    \caption{Success rate of Qwen3.5 on each project (Hands-Off: 211 of \tasks tasks; Hands-On: 383 of \tasks tasks).}
    \label{tab:qwen}
    \tablestyle
    \begin{tabular}{lrrrrrrrrrr}
    \toprule
     & AL & AC & IR & MA & NO & NR & OS & ST & VE & All \\
    \#Tasks & 104 & 63 & 118 & 89 & 29 & 204 & 157 & 63 & 22 & \tasks \\
    \midrule
    Hands-Off & 40\% & 11\% & 34\% & 31\% & 38\% & 20\% & 19\% & 8\%  & 32\% & 25\% \\
    Hands-On  & 59\% & 19\% & 46\% & 67\% & 83\% & 38\% & 41\% & 27\% & 59\% & 45\% \\
    \bottomrule
    \end{tabular}
\end{table}

As shown in Table~\ref{tab:qwen}, Qwen3.5 lies between \gmini and \gfive in both settings (17\%, 25\%, and 50\% in Hands-Off; 41\%, 45\%, and 56\% in Hands-On), far below the strongest proprietary models. Its two hardest projects, Storage (ST) and Anvil Controller (AC), are also among the hardest projects for the other models. Like other less capable models, Qwen3.5 benefits a lot from the customized Hands-On agent, which improves its success rate by 20 percentage points. If we price its tokens at the public serving rate of the model (\$0.385 and \$2.45 per million input and output tokens), Hands-Off Qwen3.5 costs about \$0.7 per task, roughly 20$\times$ cheaper than \opusfourfive, but with a much lower success rate.

We also learned one practical lesson. \tool sends long, blocking requests to the model. When we first ran Hands-On on a server shared with Hands-Off jobs, many of these requests timed out, and the success rate dropped sharply. The numbers above come from runs where Hands-On used the server alone. Users hosting their own models should check serving latency before comparing agents.

\subsection{Statistical uncertainty}
\label{app:stats}

\begin{figure}[h]
    \centering
    \includegraphics[width=0.85\linewidth]{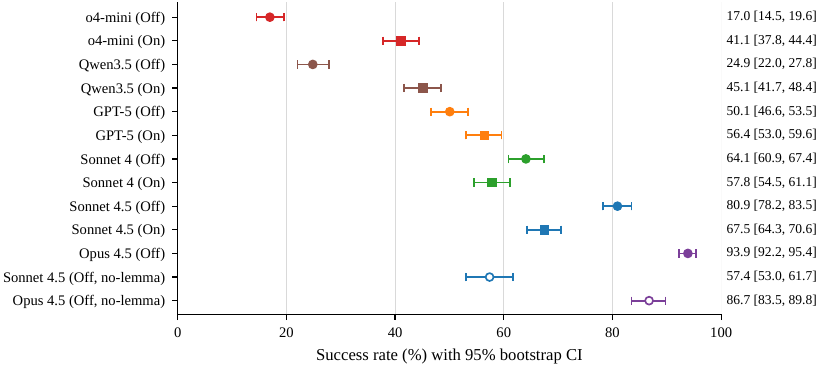}
    \caption{Success rate with 95\% bootstrap confidence interval (circles: Hands-Off; squares: Hands-On; hollow markers: no-lemma tasks).}
    \label{fig:success-ci}
\end{figure}

\MyPara{Confidence intervals.} Figure~\ref{fig:success-ci} shows the success rate of each configuration together with its 95\% bootstrap confidence interval (10{,}000 resamples of tasks). These intervals agree with the ones in Table~\ref{tab:allresults}. With \tasks tasks, the intervals are between $\pm$1.6 and $\pm$3.4 percentage points for normal tasks, and between $\pm$3.1 and $\pm$4.5 for the \noltasks no-lemma tasks.

\MyPara{Paired comparisons.} Since all configurations run on the same tasks, we compare two configurations with McNemar's test, which looks at the tasks where exactly one of them succeeds (Table~\ref{tab:mcnemar}). Every pair of neighboring models under Hands-Off differs with $p<10^{-20}$. The agent effect is significant in both directions: Hands-On helps \gmini and \gfive, and hurts \sfour and \sfourfive.

\begin{table}[h]
    \caption{Exact McNemar's tests between configurations. $b$ is the number of tasks proved by A but not B; $c$ is the number of tasks proved by B but not A.}
    \label{tab:mcnemar}
    \tablestyle
    \begin{tabular}{llrrr}
    \toprule
    A & B & $b$ (A only) & $c$ (B only) & exact $p$ \\
    \midrule
    \gmini Off      & \gfive Off        & 23  & 304 & $9.5\times10^{-64}$ \\
    \gfive Off      & \sfour Off        & 27  & 146 & $5.9\times10^{-21}$ \\
    \sfour Off      & \sfourfive Off    & 6   & 149 & $8.0\times10^{-37}$ \\
    \sfourfive Off  & \opusfourfive Off & 10  & 120 & $4.3\times10^{-25}$ \\
    \midrule
    \gmini Off      & \gmini On         & 45  & 250 & $1.3\times10^{-35}$ \\
    \gfive Off      & \gfive On         & 83  & 137 & $3.3\times10^{-4}$ \\
    \sfour Off      & \sfour On         & 108 & 55  & $4.0\times10^{-5}$ \\
    \sfourfive Off  & \sfourfive On     & 142 & 28  & $1.5\times10^{-19}$ \\
    \midrule
    \sfourfive no-lemma & \opusfourfive no-lemma & 2 & 137 & $2.8\times10^{-38}$ \\
    \bottomrule
    \end{tabular}
\end{table}

\MyPara{Run-to-run variance.} As discussed in ``Multi-run reliability'' above, repeated runs change the success rate of a single run by only 1--2 percentage points (41\%, 39\%, and 39\% for Hands-On \gmini), much smaller than the 13--33 point gaps between neighboring models. The ranking of models never changes across our repeated runs.

\MyPara{Construction of \benchhard.} \benchhard combines two groups of tasks (Table~\ref{tab:hard}). The first group contains the 37 normal tasks that none of the nine model/agent configurations in Table~\ref{tab:allresults} could prove. Recall that a normal task keeps the signatures (but not the bodies) of the helper lemmas used in the human proof, if there are any: 30 of these 37 tasks come with such lemma stubs, and the other 7 come from human proofs that use no helper lemmas. The second group contains all 58 no-lemma tasks from Anvil Controller (AC), the hardest project; \opusfourfive proves 24 of them, and no other configuration proves more. The 9 AC tasks in the first group also appear in the second group as no-lemma versions, so the 95 tasks cover 86 distinct functions. Since each configuration was run only once, a repeated run might prove a few tasks in the first group.

\begin{table}[h]
    \caption{Number of tasks in \benchhard from each project. The 9 normal AC tasks also appear among the no-lemma AC tasks.}
    \label{tab:hard}
    \tablestyle
    \begin{tabular}{lrrrrrrrrrr}
    \toprule
     & AL & AC & IR & MA & NO & NR & OS & ST & VE & All \\
    \midrule
    Normal tasks no configuration proves & 0 & 9 & 3 & 2 & 0 & 14 & 8 & 1 & 0 & 37 \\
    \quad with helper-lemma stubs        & 0 & 9 & 3 & 1 & 0 & 13 & 3 & 1 & 0 & 30 \\
    \quad without helper lemmas          & 0 & 0 & 0 & 1 & 0 & 1  & 5 & 0 & 0 & 7  \\
    \midrule
    No-lemma AC tasks                    & -- & 58 & -- & -- & -- & -- & -- & -- & -- & 58 \\
    \midrule
    \benchhard                           & 0 & 67 & 3 & 2 & 0 & 14 & 8 & 1 & 0 & 95 \\
    \bottomrule
    \end{tabular}
\end{table}

\section{Licenses and Provenance}
\label{app:license}

Table~\ref{tab:licenses} lists the repository and license of each project that \bench tasks are extracted from; we checked each license by reading the \codeIn{LICENSE} file in the repository. The license of Memory Allocator covers both the Verus contributors and the upstream mimalloc code, and NRKernel can be used under either MIT or Apache-2.0. In our release, the tasks of each project come with the project's \codeIn{LICENSE} file, and every task file keeps its copyright notice.

\begin{table}[h]
    \caption{Upstream repositories and licenses of the projects in \bench.}
    \label{tab:licenses}
    \tablestyle
    \begin{tabular}{lll}
    \toprule
    Project (task code) & GitHub repository & License \\
    \midrule
    Anvil (AL, AC)          & \texttt{anvil-verifier/anvil}                  & MIT \\
    IronKV (IR)             & \texttt{verus-lang/verified-ironkv}            & MIT \\
    Memory Allocator (MA)   & \texttt{verus-lang/verified-memory-allocator}  & MIT \\
    Node Replication (NO)   & \texttt{verus-lang/verified-node-replication}  & MIT \\
    NRKernel (NR)           & \texttt{matthias-brun/verified-nrkernel}       & MIT or Apache-2.0 \\
    Atmosphere (OS)         & \texttt{mars-research/atmosphere}              & MIT \\
    Storage (ST)            & \texttt{microsoft/verified-storage}            & MIT \\
    Vest (VE)               & \texttt{secure-foundations/vest}               & MIT \\
    \bottomrule
    \end{tabular}
\end{table}

\MyPara{Atmosphere.} We received the Atmosphere code from its authors in late November 2025, before it was public, so that we could use it to check for memorization. Atmosphere has since been released at \codeIn{github.com/mars-research/atmosphere} under the MIT license, which allows redistributing derived files as long as the notices are kept, and the code we extracted tasks from is included in this public release. In addition, the Atmosphere authors gave us written permission to redistribute the task files extracted from their code; the signed statement is included in our artifact repository.

\MyPara{Other assets.} \tool builds on AutoVerus~\cite{autoverus} (MIT), and all experiments use Verus~\cite{verussosp24,verus-oopsla} (MIT). The weights of Qwen3.5~\cite{qwen35} are released under Apache-2.0. We release \bench, the \tool agent, and all model trajectories under the MIT license.

\end{document}